\documentclass[fleqn,usenatbib]{mnras}

\usepackage{newtxtext,newtxmath}

\usepackage[T1]{fontenc}

\DeclareRobustCommand{\VAN}[3]{#2}
\let\VANthebibliography\thebibliography
\def\thebibliography{\DeclareRobustCommand{\VAN}[3]{##3}\VANthebibliography}

\usepackage{graphicx}	
\usepackage{amsmath}	
\usepackage{multicol}        
\usepackage{bm}		
\usepackage{pdflscape}	
\usepackage{longtable}  
\usepackage{caption} 
\usepackage{subfigure} 
\usepackage[flushleft]{threeparttable}
\usepackage{multirow}   
\usepackage{xfrac}   
\usepackage{xcolor} 
\usepackage{todonotes} 
\usepackage{outlines} 
\usepackage{ulem} 

\newcommand{\ignore}[1]{}

\title[MeerTRAP transients]{Discovery of 30 Galactic radio transient pulsars with MeerTRAP}

\author[J. Tian et al.]{J. Tian$^{1}$\thanks{E-mail: jun.tian@manchester.ac.uk}, S. Singh$^{1}$\thanks{E-mail: shubham.singh@manchester.ac.uk}, B. W. Stappers$^1$, J. D. Turner$^{1}$, K. M. Rajwade$^2$, M. C. Bezuidenhout$^{3,4}$, 
\newauthor M. Caleb$^{5,6}$, I. Pastor-Marazuela$^{7,1}$, F. Jankowski$^8$, V. Gupta$^{9}$, C. Flynn$^{9}$, R. Karuppusamy$^{10}$,
\newauthor E. D. Barr$^{10}$, M. Kramer$^{10}$, R. Breton$^{1}$, C. J. Clark$^{11,12}$, D. J. Champion$^{10}$, T. Thongmeearkom$^{1,13}$\\
$^{1}$ Jodrell Bank Centre for Astrophysics, Department of Physics and Astronomy, The University of Manchester, Manchester M13 9PL, UK\\
$^2$Astrophysics, The University of Oxford, Denys Wilkinson Building, Keble Road,
Oxford OX1 3RH, UK\\
$^3$Centre for Space Research, North-West University, Potchefstroom 2531, South Africa\\
$^{4}$Department of Mathematical Sciences, University of South Africa, Cnr Christiaan de Wet Rd and Pioneer Avenue, Florida Park, 1709, Roodepoort, South Africa\\
$^5$Sydney Institute for Astronomy, School of Physics, The University of Sydney, NSW 2006, Australia\\
$^6$ARC Centre of Excellence for Gravitational Wave Discovery (OzGrav), Hawthorn, 3122, Victoria, Australia\\
$^{7}$ ASTRON, the Netherlands Institute for Radio Astronomy, Oude Hoogeveensedijk 4,7991 PD Dwingeloo, The Netherlands\\
$^8$LPC2E, OSUC, Univ Orleans, CNRS, CNES, Observatoire de Paris, F-45071 Orleans, France\\
$^9$Centre for Astrophysics and Supercomputing, Swinburne University of Technology, Mail H30, PO Box 218, Hawthorn, VIC 3122, Australia\\
$^{10}$Max-Planck-Institut fur Radioastronomie, 53121 Bonn, Germany\\
$^{11}$Max Planck Institute for Gravitational Physics (Albert Einstein Institute), D-30167 Hannover, Germany\\
$^{12}$Leibniz Universität Hannover, D-30167 Hannover, Germany\\
$^{13}$National Astronomical Research Institute of Thailand, Don Kaeo, Mae Rim, Chiang Mai 50180, Thailand
}

\date{Accepted XXX. Received YYY; in original form ZZZ}

\pubyear{2025}

\begin{document}
\label{firstpage}
\pagerange{\pageref{firstpage}--\pageref{lastpage}}
\maketitle

\begin{abstract}
We present the discovery of 30 new Galactic sources from the MeerTRAP project, a commensal fast radio transient search programme using the MeerKAT telescope. These sources were all identified via a single pulse search. Most of them are likely to be rotating radio transients (RRATs) given their low pulse rates. Using data captured in our transient buffer we have localised nine sources in the image domain to arcsecond precision. This facilitates the timing of these sources and further follow-up with other telescopes. Using the arrival times of single pulses, we have constrained the periods of 14 sources, ranging from 121\,ms to 7.623\,s, and derived a phase-coherent timing solution for one of them. Follow-up observations of the MeerTRAP sources (including those published previously) performed with the Effelsberg telescope have detected regular but faint emission from three sources, confirming their long rotation period, including PSR J2218$+$2902 with a period of 17.5\,s, the fourth slowest in the radio pulsar population. A few of the sources exhibit interesting emission features, such as periodic microstructure in PSR J1243$-$0435 and possible nulling in PSR J1911$-$2020 and PSR J1243$-$0435. We find that the duty cycles of the three newly discovered pulsars are very low and follow the general trend for the duty cycle with period of known pulsars.
\end{abstract}




\section{Introduction}\label{intro}
Fast radio transients are short-duration radio signals of astrophysical origin \citep{FRT_cordes}. Such short timescale signals are thought to be associated with extreme and dynamic astrophysical phenomena because of the rapid release of immense amounts of energy from a small region \citep{FRT_and_extremephysics}. 
The fast radio transients of extragalactic origin are known as Fast Radio Bursts (FRBs, see \citealt{Lorimer24} for an overview). The origin of FRBs is still unknown \citep{Platts19, Zhang23}. 
The Galactic fast radio transients are thought to mostly originate from highly magnetized and rotating neutron stars. This includes pulsars \citep[e.g.][]{Pulsar_review_beskin}, magnetars \citep[e.g.][]{magnetar_Review_mereghetti}, and rotating radio transients (RRATs; \citealt{McLaughlin06, RRAT_review}).

RRATs are Galactic radio transients with underlying periodicity similar to pulsars but have low pulse rates \citep{RRAT_review}. Their low pulse rates make it difficult to find them in regular pulsar surveys with a periodicity search \citep{SP_vs_Periodicity_Maura, RRAT_review}. While some RRATs have long periods and high magnetic fields, similar to magnetars \citep{McLaughlin09, Lyne09}, others show emission properties similar to radio pulsars. \citet{RRAT_polarization} show that the folded profiles of RRATs are highly linearly polarized ($\sim 40\%$ on average), while their individual pulses can be up to $\sim 100\%$ linearly polarized. RRAT J0139+3336 has recently been shown to follow the rotating vector model usually seen in pulsars \citep{RRAT_RVM}. Similarly, RRAT J1854+0306 was seen in two different emission modes at two different observing epochs \citep{RRAT_modes}. Microstructure has been seen in at least two RRATs (J0139+3339 and J1918-0449) with the Five-hundred-meter Aperture Spherical Telescope (FAST; \citealt{RRAT_micro, RRAT_RVM}). These observed features resemble those seen in pulsar radio emission. 
Therefore, studying RRATs will help us understand the link between RRATs and normal pulsars.

Magnetars are another class of Galactic transients discovered through their high energy transient behaviour, but a subclass of these sources show radio emission as well (see \citealt{magnetar_Review_mereghetti} and \citealt{magnetar_review_kaspi} for an overview). RRATs and radio-loud magnetars have some common properties such as their sporadic radio emission and similar pulse morphology, including microstructure \citep{microstrucutre_kramer}. Moreover, it has been proposed that some RRATs could be associated with X-ray outbursts of magnetars \citep[e.g.][]{Rea09, Archibald17}, which is supported by the similar spin-down properties of RRATs and magnetars as well as their typically strong magnetic fields (up to $5\times10^{13}$\,G; \citealt{McLaughlin09, Lyne09}).

There is evidence of a connection between FRBs and magnetars. An FRB-like signal from the Galactic magnetar SGR 1935+2154 \citep{CHIME20, Bochenek20} suggests that at least some FRBs could arise from magnetars. This emission also has luminosity comparable to the faintest and nearest extragalactic FRBs \citep{Patel18, CHIME19b}. In addition, a few FRBs show pulsar-like polarisation position angle swings that are in agreement with the rotating vector model \citep{Mckinven25, Liu25}, and a few other FRBs show quasi-periodic sub-structure \citep{Majid21, CHIME22, Pastor-Marazuela23}. Given these observational similaries between RRATs, magnetars, and FRBs, finding and studying more Galactic transients is crucial to advance our understanding of their connection.

The slow rotating neutron star population is becoming increasingly interesting with new discoveries of long-period (period of a few tens of seconds, e.g., \citealt{12s_pulsar, 23s_pulsar, 76s_pulsar, WangY25}) and ultra-long period (ULP; period of a few minute timescales, e.g., \citealt{18m_source,Hurley24}) sources. 
Not all of them may be neutron stars, and some ULPs are likely to be white dwarf binaries, e.g. GLEAM-X J0704$-$37 \citep{Hurley24} and ILT J1101+5521 \citep{Ruiter25}.
If some of these sources are indeed neutron stars, it will
challenge the conventional understanding of radio emission from neutron stars \citep{76s_pulsar}. There is also tentative evidence suggesting that long-period radio sources could be radio-loud magnetars \citep{76s_pulsar, Surnis23, Lee25}. There are only half a dozen sources with periods between a few tens of seconds and a few minutes \citep{23s_pulsar, 12s_pulsar, 76s_pulsar, Surnis23, Dong25, WangY25}. Galactic transients like RRATs and magnetars are mostly long-period sources. Discovery and characterization of new Galactic transients can add more sources to the slow-rotating neutron star population or reveal different source populations.

There are two commonly used methods to search for short-timescale radio signals: periodicity searches and single pulse searches. Fast transients such as FRBs that do not have any strict periodicity can only be discovered by a single pulse search \citep{FRBsearch_review}. Transients with an underlying periodicity can be discovered by periodicity search methods \citep{staelin, Ransom_2002, FFA_morello}. Even when Galactic transients like RRATs do have underlying periodicity, they are usually discovered in single pulse searches due to their sporadic emission \citep{RRAT_review, SP_vs_Periodicity_Maura}.
Some RRATs also show pulsar-like, faint emission \citep[e.g.][]{Zhang24, Mcsweeney25}, which could be detected via a periodicity search.

There have been many single pulse searches for Galactic transient sources, e.g. realfast at the Jansky Very Large Array \citep{Law18}, CRACO at the Australian Square Kilometre Array Pathfinder \citep{WangZ25}, GHRSS and SPOTLIGHT at the Giant Metrewave Radio Telescope \citep{Bhattacharyya19, Panda24}, HTRU-South at the Murriyang telescope \citep{Keith10}, HTRU-North at the Effelsberg telescope \citep{Barr13}, and SMIRF at the UTMOST telescope \citep{Venkatraman20}.
\citet{Good21} discovered seven new Galactic sources with CHIME/FRB \citep{CHIME18}, of which four appear to be RRATs and the others intermittent or nulling pulsars, which display intermittent emission but behave like normal pulsars when active. This is followed by another set of discoveries by CHIME/FRB using the same method, which includes 14 RRATs and seven pulsars \citep{Dong23}. The FAST Galactic Plane Pulsar Snapshot Survey, in addition to a traditional periodicity search, has also performed a single pulse search and discovered 76 new transient radio sources \citep{Zhou23}. 

We have been using the Meer(more) TRAnsients and Pulsars (MeerTRAP; \citealt{Sanidas18, Bezuidenhout22, Rajwade22, Caleb23, Jankowski23, Driessen24}) project to perform commensal, time-domain searches for fast transients in real-time with the MeerKAT telescope in South Africa. Thanks to the wide field of view and high sensitivity of MeerKAT, this search program has discovered a few dozen Galactic radio transients \citep{Bezuidenhout22, MeerTRAP_james}. \citet{Bezuidenhout22} presented the first discovery of 12 Galactic sources by MeerTRAP, followed by the report of another 26 new sources by \citet{MeerTRAP_james}. In this paper, we present 30 more MeerTRAP Galactic sources. In Section~\ref{sec:obs}, we briefly summarise the observational configuration of MeerKAT and the real-time search pipeline, and describe the offline data processing. We also introduce follow-up observations including sources previously discovered. Our results are then presented in Section~\ref{sec:results}. We discuss some interesting sources and future prospects in Section~\ref{sec:disc}, followed by conclusions in Section~\ref{sec:concl}.

\section{Observations and Data Reduction}\label{sec:obs}
\subsection{MeerKAT data processing}
\subsubsection{Array configuration and data recording}
MeerTRAP is a commensal project that piggybacks on other science observations to search for fast transients 
\citep{meertrap_concept, Bezuidenhout22, MeerTRAP_james}. The sky coverage depends on the targets of these surveys and the observing frequency band is determined by the MeerKAT receivers being used during these observations \citep{Bezuidenhout22}. 
The sources reported in this paper were discovered either in the UHF (544 - 1088 MHz; \citealt{UHF_band}) or L (856 - 1712 MHz; \citealt{L_band}) bands.
The time domain data is recorded simultaneously with coherent and incoherent beam modes. The incoherent beam uses all the working antennas available at the time of observation, which can be up to 64.
The MeerKAT incoherent beam has very wide field of view (half power area of 1 deg$^2$ in the L band and 2.4 deg$^2$ in the UHF band; \citealt{Mauch20}). Up to 768 coherent beams are formed to partially cover this field of view. The coherent beams only use the central 40--44 antennas within a baseline of 1 km. 
These coherent beams are arranged so that they are always between one and five times more sensitive than the incoherent beam \citep{Rajwade22}. Data from these beams are recorded at a time resolution of 482 $\mu$s in the case of UHF band or 306 $\mu$s in the case of L-band observations. The number of frequency channels is 1024 for these observations. 

\subsubsection{Real-time pulse searches}
The MeerTRAP search pipeline searches the coherent and incoherent beam data for dispersed single pulses in real time. The data are cleaned for zero-DM radio frequency interference (RFI; \citealt{Eatough09}). Then {\sc iqrm} cleaning is employed with a threshold of 3$\sigma$ to filter out narrowband RFI \citep{Morello22}. To further reject the zero-DM RFI, all the detections with DM below 20 $\mathrm{pc}$ $\mathrm{cm}^{-3}$ are rejected. We search up to DM of 2100 $\mathrm{pc}$ $\mathrm{cm}^{-3}$ in the case of UHF and up to 3600 $\mathrm{pc}$ $\mathrm{cm}^{-3}$ in the case of L-band observations. The detections above 8$\sigma$ significance are then sifted and classified using {\sc frbid}\footnote{\url{https://github.com/Zafiirah13/FRBID}}. See \citet{Rajwade22} for a more detailed description of the search pipeline.

\subsubsection{Localisation}\label{sec:image}
Since the coherent beam uses a maximum baseline of 1 km, the discoveries from the time domain search have moderate uncertainties in their location ($\sim$ arcmin). An extra localisation effort is needed to increase the accuracy of the source coordinates that then can be used for more sensitive follow-up observation of the source. The coherent beams are arranged to overlap at their 50\% sensitivity level. If a source has been detected in three or more coherent beams, it may be localised to the overlapping region of those beams assuming that the pulse detected by a single tied-array beam of MeerKAT originates within that beam, which is usually the case except for beams lying on the edge of the tiling. This is carried out using the multibeam localisation tool {\sc seekat}\footnote{\url{https://github.com/BezuidenhoutMC/SeeKAT}} \citep{Bezuidenhout23}, with the point spread function of each observation being simulated using {\sc mosaic}\footnote{\url{https://github.com/wchenastro/Mosaic}} \citep{Mosaic}.

A more accurate localisation can be achieved for detections that trigger the transient buffer to store $\sim300$ ms of voltage data from individual antennas \citep{Rajwade24}. The data in the transient buffer is correlated with \texttt{xGPU}\footnote{\texttt{xGPU}: \url{https://github.com/GPU-correlators/xGPU}} \citep{clark_accelerating_2011} and then converted to measurement sets with \texttt{DifX}~\citep{deller_difx_2007, deller_difx-2_2011}. 
Then on-pulse and off-pulse images are made with {\sc wsclean} \citep{Offringa14}. The transient buffer stores data from all antennas in the total array, not just those used in beamforming and this can be up to 64 with a maximum baseline of $\sim8$\,km \citep{Jonas16}. 
The pulse S/N is therefore expected to be higher in these data than in the coherent beams.
The on-pulse and off-pulse images are then compared to identify the transient point source. The final position is determined using the Python Blob Detector and Source Finder ({\sc pybdsf}\footnote{\url{https://www.astron.nl/citt/pybdsf/}}), and the absolute astrometry is corrected using catalogued sources from the Rapid ASKAP Continuum Survey (RACS; \citealt{McConnell20}), as described in \citet{Driessen22, Driessen24}. This method usually results in a localisation accuracy of $\sim1$ arcsec, which consists of three components: the source fitting error given by {\sc pybdsf}, the systematic uncertainty in the RACS positions and the median offset of the positions after the astrometric correction. More details about the voltage data reduction and imaging can be found in \citet{Rajwade24}, \citet{Tian24b}, and \citet{pastor-marazuela_localisation_2025}.

\subsection{DM determination}\label{sec:DM}

For each source presented here, we select the brightest pulse detection to determine the S/N maximising DM. 
Then we refined this DM using {\sc scatfit}\footnote{\url{https://github.com/fjankowsk/scatfit}}, which models subbanded pulse profiles with exponentially modified Gaussians that take into account the pulse broadening due to scattering and intra-channel dispersion smearing \citep{Jankowski23}. 
This method is based on a Bayesian analysis by Markov Chain Monte Carlo sampling and provides a refined DM measurement and uncertainty. For some pulses with low S/Ns, we downsampled the data in time by a factor of up to 8, and in frequency to 8 subbands before fitting a scattered pulse model. Subbands with $\text{S/N}<3$ were not included in the fit. 
The DM with the lowest Bayesian information criterion (BIC) for each source is listed in Table~\ref{tab:properties}.

\subsection{Period search}\label{timing}

For some sources that have multiple pulses detected, we calculated the times of arrival (TOAs) of pulses to measure their spin period. We used {\sc make\_toas}\footnote{\url{https://bitbucket.org/meertrap-ipm/mtcutils/src/jturner-timing/}}, an implementation based on {\sc mtcutils}\footnote{\url{https://bitbucket.org/vmorello/mtcutils} by Vincent Morello, modified by In\`{e}s Pastor-Marazuela and James Turner}, to calculate the TOA of each single pulse. This was done by convolving single-Gaussian templates with dedispersed time series using {\sc spyden}\footnote{\url{https://bitbucket.org/vmorello/spyden}}. The TOA is given by the peak of the best matched Gaussian, and its uncertainty is the pulse width divided by the S/N. Under this method, pulses that have more complicated shapes or multiple components will have TOAs that are more noisy, however this noise will always be much smaller than the rotational period.

An initial period estimate is attempted based on the spacing between TOAs measured for a source. We used {\sc rratsolve}\footnote{\url{https://github.com/v-morello/rratsolve}} to find the largest period connecting all TOAs. This method requires at least 3 TOAs closely spaced in time. If a source is detected multiple times over a long time span, we can use the pulsar timing software {\sc tempo2}\footnote{\url{https://bitbucket.org/psrsoft/tempo2/src/master/}} \citep{Hobbs06} to obtain a phase-connected timing solution. The initial pulsar ephemeris contains the period estimated by {\sc rratsolve} and the position from the relevant localisation technique. Since the timing precision is limited by the number of pulses detected and their spacing in time and the localisation accuracy of the source, we are able to obtain timing solutions for only one source.

\subsection{Effelsberg Follow-up}\label{sec:followup}
A few well-localised sources were followed up with the 100-metre Effelsberg radio telescope \citep{Effelseberg_paper}. The observations were carried out with the Ultra Broad-Band (UBB) receiver on 2024 January 17 and 2024 February 17 with a frequency coverage of 1.3--6\,GHz. The entire band was split into 5 subbands: 1.3--1.9\,GHz, 1.9--2.6\,GHz, 3.0--4.1\,GHz, 4.1--5.2\,GHz and 5.2--6.0\,GHz. The observations were conducted in the search mode with 8-bit full Stokes and $64\,\mu$s time resolution and 1\,MHz frequency resolution. Each source was observed continuously for 2 hours, and a 2\,min scan of a noise diode was performed prior to each observation.\\

The data obtained from these observations were then searched for single pulses in the DM range of $\pm10\%$ of the discovery DM using {\sc transientx}\footnote{\url{https://github.com/ypmen/TransientX}} \citep{TransientX}. {\sc filtool} was used for RFI cleaning. We chose a DM step size of $0.1\,\text{pc}\,\text{cm}^{-3}$ and a maximum boxcar width of 100\,ms. Single pulse events detected with a S/N above 8 were manually inspected for real signals.
We also searched these observations with the Fast Folding Algorithm (FFA, \citealt{singh_FFA, FFA_morello, staelin}) as the MeerTRAP sources may emit weak regular periodic signals. We cleaned and dedispersed the data to the best DM using {\sc transientx}. Then we used the {\sc rseek} utility of the {\sc riptide}\footnote{\url{https://github.com/v-morello/riptide}} package \citep{FFA_morello} to search for periodic signals in the dedispersed time series. The period search range was set to $\pm 10\%$ of the nominal period reported in table~\ref{tab:properties}, and the duty-cycle range was 0.2\% to 20\%. All candidates with S/N $>8$ were folded and manually inspected. Detecting regular faint periodic signals from the MeerTRAP sources would allow us to refine the period estimated by the brute-force method {\sc rratsolve} and study their folded pulse properties.

\section{Results}\label{sec:results}

\begin{figure*}
\centering
\includegraphics[width=\textwidth]{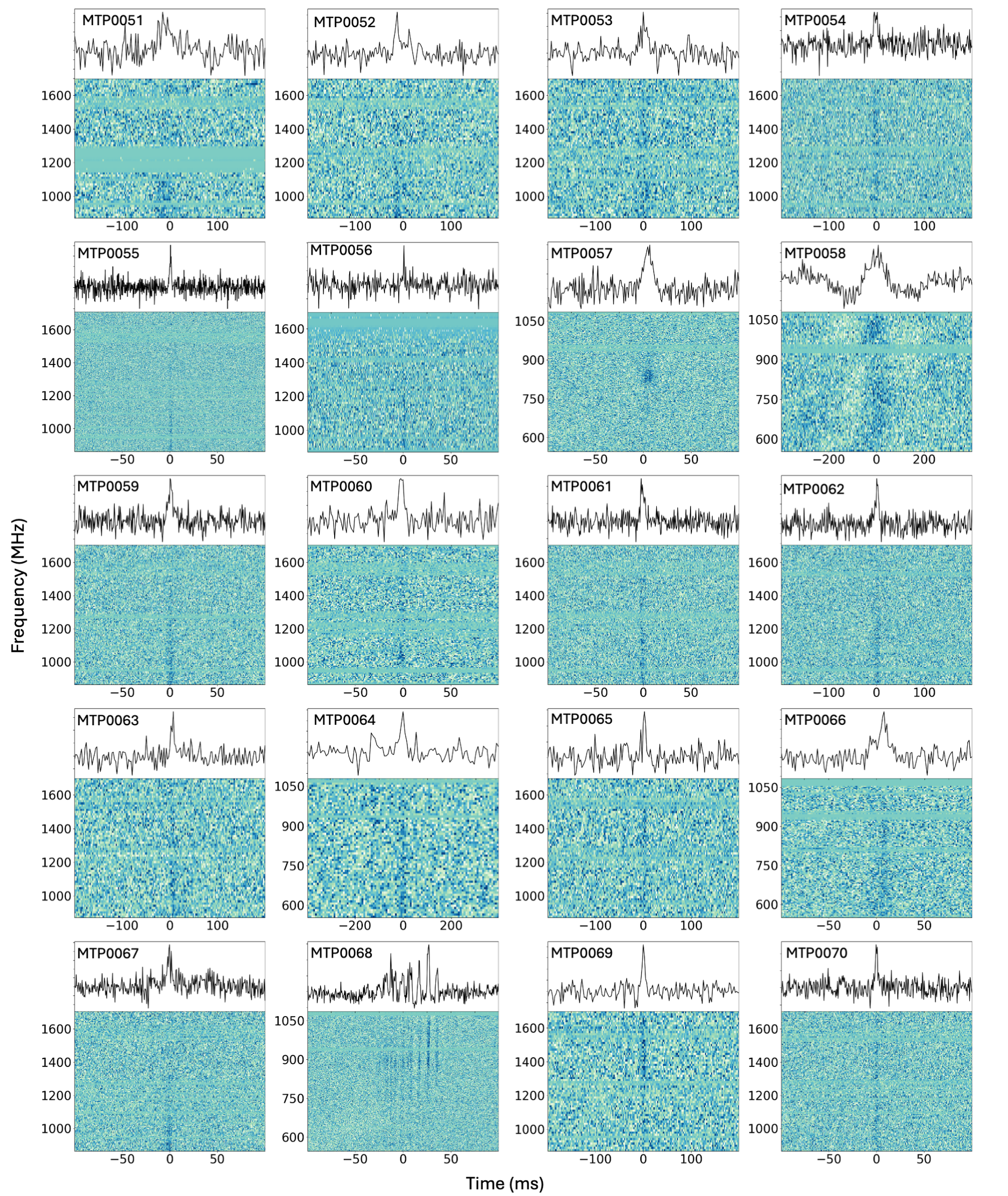}
\caption{The discovery pulses of the 30 Galactic MeerTRAP sources. Each panel shows the dynamic spectrum (bottom) and frequency-averaged pulse profile (top) from the filterbank data dedispersed to the best DM given in Table~\ref{tab:properties}. The data of faint pulses have been donwsampled in frequency and time by a factor of up to 32 and 16. The source name (see Table~\ref{tab:detections}) is given to each pulse in the top-left corner. Blank horizontal lines are either missing channels or flagged due to RFI. The plot of MTP0058 is affected by RFI and zero-DM flagging.}
\label{fig:discovery}
\end{figure*}

We present 30 new Galactic sources discovered by MeerTRAP via single pulse search. The discovery information of these sources is listed in Table~\ref{tab:detections}, and the discovery pulses are shown in Figure~\ref{fig:discovery}. We note that eight of these sources were new at the time of first detection, but were independently discovered by other surveys: 
MTP0067 was detected by the Max-Planck-Institut fur Radioastronomie (MPIfR)–MeerKAT Galactic Plane Survey (MMGPS; \citealt{Padmanabh23}; see also \citealt{Stappers18});
MTP0075 was detected by the TRAPUM UHF survey for pulsars in Fermi-LAT gamma-ray sources (Thongmeearkom et al. in preparation; see also \citealt{Clark23}).
MTP0070 and MTP0078 were detected by the FAST Galactic Plane Pulsar Snapshot (GPPS) survey \citep{Han25}; MTP0074 was detected by CHIME/FRB\footnote{The source is listed in the catalogue: \url{http://catalog.chime-frb.ca/galactic}}; MTP0054 was detected in the archival data of the Parkes multibeam pulsar survey \citep{Sengar23}; MTP0068 was detected by an UTMOST \citep{Gupta2022} pulsar and FRB survey (see Section \ref{subsec:MTP0068_UTMOST_detection}); MTP0055 was detected by the Southern-sky MWA Rapid Two-meter (SMART; \citealt{Bhat23b, Bhat23a}) pulsar survey\footnote{The source is listed in the catalogue: \url{https://mwatelescope.atlassian.net/wiki/spaces/MP/pages/24970773/SMART+survey+candidates}}. We highlight the eight sources in Table~\ref{tab:detections}. Although these sources have been reported, we measured the properties such as the location, period, and fluence for them based on the MeerKAT data (see Table~\ref{tab:properties}).


\begin{table*}
\centering
\caption{Discovery information for the reported transient sources. We highlight the sources independently discovered by other surveys with the asterisk ($\ast$). See Section~\ref{sec:results} for more details.} 
\label{tab:detections}
\begin{tabular}{llcclllrrrrl}
\hline
MTP & PSR J2000 & Disc. MJD & Disc. project$^{1}$ & Disc. & Disc. &  Disc. DM & $T_{\text{obs}_\text{IB}}^{3}$ & $T_{\text{obs}_\text{CB}}^{3}$ & $N_{\text{det}}^{4}$ & $N_{\text{e}}^{5}$ & Band$^{6}$ \\
name &name&&&mode$^{2}$&S/N& (pc cm$^{-3}$)&(hr)&(hr)&&\\
\hline
MTP0051 &  & 59627.859034 & SCI-20200703-MK-01 & IB & 8.8 & 99.16 & 92.1 & 0.32 & 2 & 2 & L \\
MTP0052 &  & 59622.359342 & SCI-20200703-MK-01 & CB & 8.9 & 294.41 & 27.06 & 0.16 & 2 & 1 & L \\
MTP0053 &  & 59637.706047 & SCI-20200703-MK-01 & CB & 9.4 & 301.47 & 31.55 & 0.23 & 1 & 1 & L \\
MTP0054$^{\ast}$ & J1655-40  & 59640.062581 & SCI-20200703-MK-01 & CB & 10.6 & 89.95 & 48.62 & 0.16 & 1 & 1 & L \\
MTP0055$^\ast$ & J0917-4420 & 59636.753994 & SCI-20200703-MK-01 & CB & 12.1 & 46.36 & 14.39 & 1.91 & 26 & 8 & L,UHF \\
MTP0056 &  & 59657.066965 & SCI-20180516-PW-01 & CB & 8.2 & 85.35 & 56.73 & 16.99 & 1 & 1 & L \\
MTP0057 & J2317-4746 & 59653.460728 & SCI-20180516-NG-02 & CB & 9.9 & 23.64 & 26.33 & 1.93 & 6 & 1 & UHF \\
MTP0058 & J2218-1229 & 59673.167861 & SCI-20180516-NG-02 & CB & 9.5 & 29.78 & 2.86 & 2.44 & 6 & 1 & UHF \\
MTP0059 &  & 59699.528926 & SCI-20200703-MK-01 & CB & 11.9 & 151.04 & 34.2 & 0.23 & 1 & 1 & L \\
MTP0060 & J1953-6111 & 59702.000543 & SCI-20180516-EB-01 & IB & 9.3 & 41.45 & 63.17 & 1 & 60 & 5 & L \\
MTP0061 &  & 59707.893381 & SCI-20200703-MK-01 & CB & 12 & 35.30 & 42.1 & 
0.16 & 2 & 1 & L \\
MTP0062 &  & 59718.143078 & SCI-20180516-MB-02 & IB & 10.6 & 46.05 & 70.29 & 0.73 & 1 & 1 & L \\
MTP0063 & J1817-1932 & 59738.97037 & SCI-20180516-PW-01 & CB & 8.9 & 217.05 & 23.66 & 1.24 & 128 & 5 & L \\
MTP0064 &  & 59744.751323 & SCI-20180516-NG-02 & CB & 8.5 & 65.08 & 12.37 & 0.94 & 1 & 1 & UHF \\
MTP0065 & J1748-3616 & 59733.922209 & SCI-20180516-PW-01 & CB & 8.4 & 268.01 & 71.44 & 3.92 & 9 & 4 & L \\
MTP0066 &  & 59730.319017 & SCI-20180923-MK-02 & IB & 10.9 & 149.20 & 2.23 & 1.58 & 2 & 2 & UHF \\
MTP0067$^\ast$ & J0933-4604 & 59782.389429 & SCI-20200703-MK-01 & CB & 11.6 & 120.34 & 35.38 & 0.17 & 33 & 2 & L \\
MTP0068$^\ast$ & J1243-0435 & 59823.507436 & SCI-20180516-NG-02 & CB & 9.1 & 21.49 & 3.9 & 2.16 & 5 & 1 & UHF \\
MTP0069 & J1548-5229 & 59824.761269 & SCI-20180516-MB-02 & CB & 10.8 & 366.25 & 74.8 & 0.5 & 3 & 2 & L \\
MTP0070$^\ast$ & J1909+0310g & 59833.705494 & SCI-20180516-MB-02 & CB & 10.8 & 110.83 & 25.8 & 1.99 & 1 & 1 & L \\
MTP0071 &  & 59835.497437 & SCI-20200703-MK-02 & IB & 12.0 & 98.55 & 27.14 & 2.29 & 1 & 1 & UHF \\
MTP0072 & J1831-1141 & 59862.615789 & SCI-20180516-MB-02 & CB & 8.9 & 40.83 & 61.25 & 0.04 & 4 & 3 & L \\
MTP0073 &  & 59852.504214 & SCI-20180516-PW-01 & CB & 8.3 & 304.54 & 110.97 & 0.42 & 1 & 1 & L \\
MTP0074$^\ast$ & J1956+3544 & 59883.607779 & SCI-20180516-PW-01 & IB & 11.4 & 153.81 & 1.74 & 0 & 16 & 3 & L,UHF \\
MTP0075$^\ast$ & J1303-4713 & 59882.153474 & SCI-20180923-MK-02 & CB & 14.4 & 82.89 & 26.88 & 0.66 & 3 & 2 & UHF \\
MTP0076 &  & 59884.557832 & SCI-20200703-MK-01 & CB & 10.1 & 283.05 & 30.64 & 0.34 & 3 & 2 & L \\
MTP0077 & J1843-0858 & 59921.378900 & SCI-20180516-MB-02 & CB & 9.9 & 300.86 & 74.79 & 3.68 & 11 & 4 & L,UHF \\
MTP0078$^\ast$ & J1914+0217 & 59897.552995 & SCI-20180516-MB-02 & CB & 16.0 & 161.18 & 18.92 & 1.77 & 12 & 7 & L \\
MTP0079 & J1816-2419 & 59964.254553 & SSV-20220221-SA-01 & CB & 10.7 & 269.24 & 49.43 & 9.3 & 34 & 6 & L \\
MTP0080 & J1848+0009 & 59846.768047 & SCI-20180516-PW-01 & IB & 8.4 & 392.23 & 104.4 & 3.74 & 6 & 2 & L \\
\hline
\end{tabular}
\begin{tablenotes}
    \item 1: The proposal id of the discovery project for each source.
    \item 2: The observing mode of the discovery observation for each source with "CB" and "IB" denoting coherent and incoherent beam, respectively.
    \item 3: The total observing time in the IB and CB mode.
    \item 4: The total number of pulses detected from each source.
    \item 5: The number of epochs when the source was detected.
    \item 6: The observing band at which each source has been detected.
\end{tablenotes}
\end{table*}

\renewcommand{\thefigure}{\arabic{figure} (Continued.)}
\addtocounter{figure}{-1}

\begin{figure*}
\centering
\includegraphics[width=\textwidth]{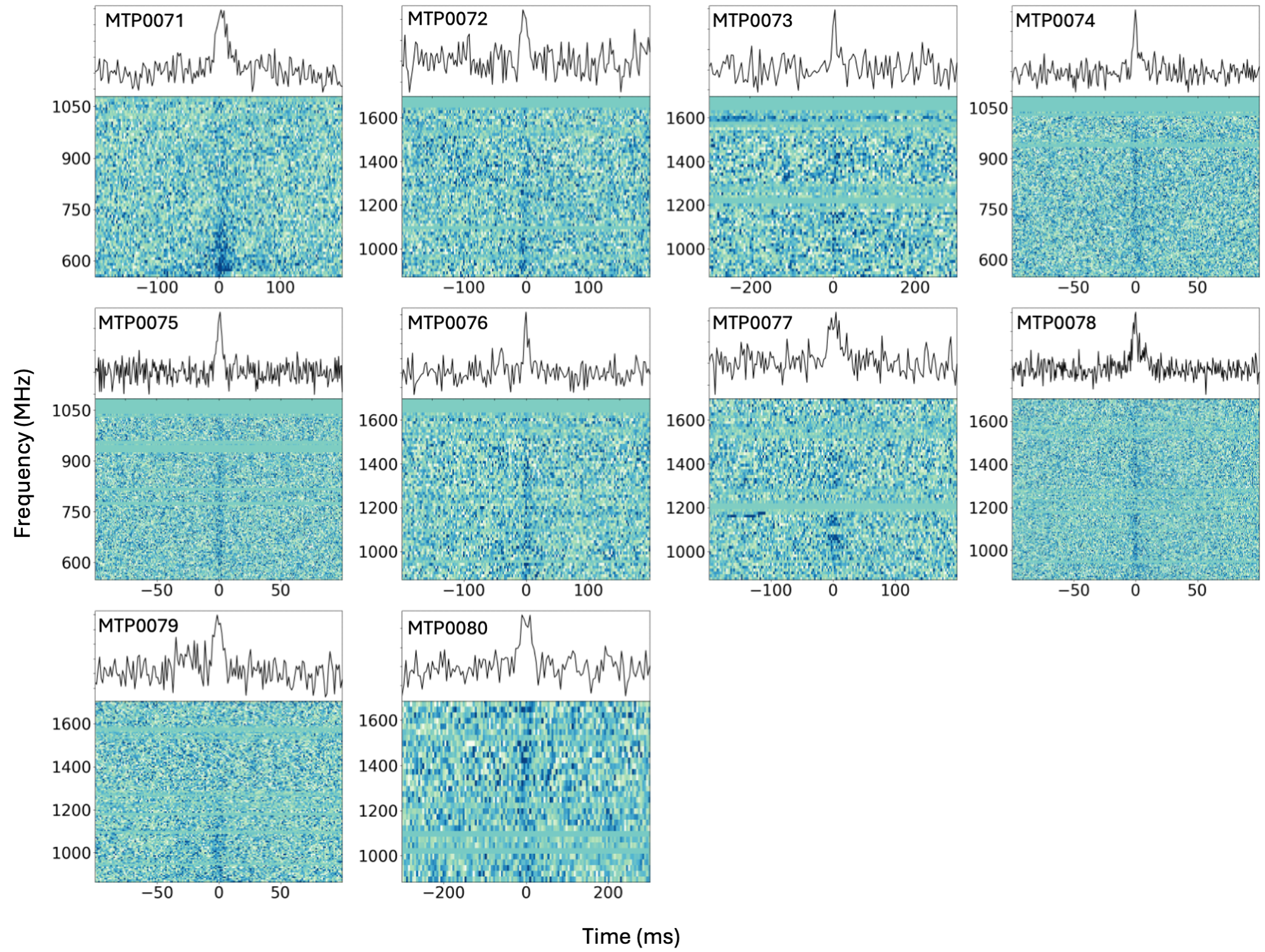}
\caption{}
\end{figure*}

\renewcommand{\thefigure}{\arabic{figure}}

\begin{table*}
\centering
\caption{Localisation, DM, distance, fluence and period measured for each transient source. Again, the sources independently discovered by other surveys are highlighted with the asterisk ($\ast$). We provide positional uncertainties for the sources localised with imaging or beam overlapping. For the other sources, their positional errors can be estimated by the beam size of the respective observing mode, i.e. $\sim1$\,arcmin for the coherent beam and $\sim1$\,deg for the incoherent beam. We do not include the distances of MTP0051, MTP0062, MTP0066 and MTP0071 because they have been detected only in incoherent beams and have large positional errors (see Section~\ref{sec:positions}).
We provide a range of fluences measured for individual pulses for the sources with imaging localisation and a lower limit for the other sources as we do not know the source location in the beam. We provide period estimates for the sources with multiple pulse detections. For the sources with only two pulse detections, we set the separation between the two pulses as the upper limit on their period.
}
\label{tab:properties}
\begin{tabular}{lllllllll}
\hline
MTP name & RA & Dec & Method$^{1}$ & DM & $D_\text{<NE2001/YMW16>}$ & Fluence & P &  P Epoch  \\
&(hh:mm:ss)&(dd:mm:ss)& & (pc cm$^{-3}$)& (kpc) & (Jy ms) & (s)&(MJD)\\
\hline
MTP0051 & 10:46:46 & -59:19:26 & IB & 101.1(7) & \ignore{2.3/1.5} & $\geq0.30-0.59$ &  &   \\
MTP0052 & 16:37:09 & -53:26:11 & CB & 296(2) & 6.4/13.7 & $\geq0.26-0.41$ & & \\
MTP0053 & 8:44:31 & -47:44:50 & CB & 302(1) & 0.6/7.0 & $\geq0.36$ & & \\
MTP0054$^{\ast}$ & 16:55:43 & -40:49:09 & CB & 92(1) & 1.8/2.5 & $\geq0.28$ & &  \\
MTP0055$^{\ast}$ & 09:17:52.9(1) & -44:20:53(1) & Imaging & 45.8(1) & 0.5/0.1 & $0.28-1.74$ & 2.58071(3) & 59835.500322 \\
MTP0056 & 15:45:43 & -47:43:50 & CB & 86(2) & 3.0/2.8 & $\geq0.09$ &  & \\
MTP0057 & 23:17:28(2) & -47:46:53(25) & Imaging & 15.9(3) & 0.7/1.8 & $\geq0.32-0.77$ & 1.7331602(8) & 59653.481429 \\
MTP0058 & 22:18:44 & -12:29:13 & CB & 26.8(6) & 1.3/>25 & $\geq1.36-2.28$ & 0.1627363(1) & 59673.192548 \\
MTP0059 & 9:03:34 & -50:16:50 & CB & 150.9(3) & 1.9/0.6 & $\geq0.25$ & &  \\
MTP0060 & 19:53:07(2) & -61:12:01(7) & Imaging & 43.0(1) & 1.6/4.2 & $\geq0.11-0.46$ & 0.46071448(3) & 60607.913720 \\
MTP0061 & 16:47:30 & -41:17:18 & CB & 35.7(3) & 1.0/0.9 & $\geq0.24-0.27$ & & \\
MTP0062 & 18:30:48 & -10:59:28 & IB & 45.6(5) & \ignore{1.4/1.0} & $\geq0.24$ & & \\
MTP0063 & 18:17:12.6(1) & -19:32:48(2) & Imaging & 214.5(2) & 4.0/3.9 & $0.15-1.29$ & 1.22912(2) & 59769.876545 \\
MTP0064 & 16:54:02 & -0:23:10 & CB & 65.4(6) & 3.8/>25 & $\geq0.70$ &  & \\
MTP0065 & 17:48:14.12 & -36:15:50.7 & CB & 266.6(5) & 5.7/14.7 & $0.31-0.77$ &  7.62348(5) & 59747.903975 \\
MTP0066 & 6:57:26 & -46:58:49 & IB & 148.4(4) & \ignore{>50/0.5} & $\geq0.48-0.62$ &  & \\
MTP0067$^{\ast}$ & 09:33:53.2(1) & -46:04:57(1) & Imaging & 120.8(1) & 1.4/0.4 & $0.15-1.08$ & 3.66981(6) & 59782.391977 \\
MTP0068$^{\ast}$ & 12:43:23(2) & -04:35:36(26) & Imaging & 12.0(1) & 0.6/1.0 & $\geq0.40-0.79$ & 4.86772(2) & 59823.512112 \\
MTP0069 & 15:48:52.8(7) & -52:29:07(6) & Imaging & 366.0(5) & 7.4/6.2 & $0.16-0.41$ & $\leq4.85$ & 60646.374716 \\
MTP0070$^{\ast}$ & 19:09:49 & +3:10:34 & CB & 110.7(4) & 3.7/4.2 & $\geq0.18$ &  & \\
MTP0071 & 9:17:52 & -44:13:20 & IB & 98.1(2) & \ignore{0.6/0.4} & $\geq1.04$ & & \\
MTP0072 & 18:31:05.1(1) & -11:41:18(1) & Imaging & 46.1(2) & 1.4/1.1 & $0.37-0.70$ & & \\
MTP0073 & 10:50:17 & -62:22:10 & CB & 304(1) & 6.0/2.5 & $\geq0.23$ & & \\
MTP0074$^{\ast}$ & 19:56:28.0(1) & +35:44:24(2) & Imaging & 153.5(1) & 6.4/7.0 & $0.34-1.60$ & 0.875567(1) & 59886.691847 \\
MTP0075$^{\ast}$ & 13:03:32 & -47:13:12 & CB & 82.6(1) & 2.6/3.6 & $\geq0.30-0.44$ & & \\
MTP0076 & 16:53:01 & -37:45:21 & CB & 283(1) & 5.7/15.6 & $\geq0.23-0.33$ &  & \\
MTP0077 & 18:40:33 & -8:09:03 & CB & 300.0(4) & 5.2/5.0 & $\geq0.38-1.26$ & 0.1211853(2) & 60218.696604 \\
MTP0078$^{\ast}$ & 19:14:46(2) & +2:18:13(30) & Imaging & 161(1) & 5.0/7.4 & $\geq0.19-2.01$ & $\leq2.02$ & 60292.481327 \\
MTP0079 & 18:16:39.6(1) & -24:19:01(1) & Imaging & 269.4(7) & 6.1/12.8 & $0.29-1.14$ & 4.612833(8) & 59721.920209 \\
MTP0080 & 18:48:58.3(1) & +00:09:19(1) & Imaging & 393.4(4) & 6.8/5.5 & $0.69-2.86$ & 4.70784(8) & 59972.199326 \\
\hline
\end{tabular}
\begin{tablenotes}
    \item 1: We localised the sources using different methods. For those localised only in the CB or IB, their positional errors are $\sim1$\,arcmin and $\sim1$\,deg, respectively (see Section~\ref{sec:positions}).
\end{tablenotes}
\end{table*}

\subsection{Source positions}\label{sec:positions}
Nine of the new Galactic sources have triggered the transient buffer and thus can be imaged using the channelised voltage data. We created and compared the on-pulse and off-pulse images of these sources and identified them in the on-pulse images, as shown in Figure~\ref{fig:images}. After applying the astrometric correction as described in Section~\ref{sec:image}, we localised the nine sources to $\sim$arcsec precision, as given in Table~\ref{tab:properties}. Note that the localisation of MTP0069 is significantly poorer than the other sources as the voltage data were recorded when MeerKAT was observing in the sub-array mode with a maximum baseline of $\sim1$\,km.
Another four sources with more than three CB detections, i.e. MTP0057, MTP0060, MTP0068 and MTP0078, were localised using {\sc seekat}. Their best-fit positions along with $1\sigma$ statistical uncertainties are listed in Table~\ref{tab:properties}.

MTP0065 was triggered on a faint pulse with a S/N of 8.8, and cannot be localised in the image. Therefore, we adopted the detection beam of the brightest pulse as the position of this source and the beam size as the uncertainty, as given in Table~\ref{tab:properties}.

For the other sources without voltage data or multiple CB detections, we adopted the detection beam of the brightest pulse as their position. Assuming again that the pulse detected by a single tied-array beam of MeerKAT originates within that beam, we can estimate the positional error by the beam size, i.e. $\sim1$arcmin in the L-band \citep{Bezuidenhout23}. Note that for sources detected only in incoherent beams, i.e. MTP0051, MTP0062, MTP0066 and MTP0071, their positional errors are much larger ($\sim$deg). All source positions are listed in Table~\ref{tab:properties}.

We derived the distances of the MeerTRAP sources based on the \texttt{NE2001} \citep{Cordes02} and \texttt{YMW16} \citep{YMW16} Galactic electron density models, as given in Table~\ref{tab:properties}. Note that we do not provide distances for the sources detected only in incoherent beams due to their large positional errors. As can be seen, there are large discrepancies in the distance estimates from these two models for some of the sources. Specifically, the sources located close to the Galactic anti-centre and at low Galactic latitudes are predicted to be more distant by \texttt{NE2001} than by \texttt{YMW16}. This is consistent with the findings from the comparison of the two models \citep{Price21}. Interestingly, MTP0058 and MTP0064 have DMs slightly higher than the Galactic contribution predicted by \texttt{YMW16}. However, given that \texttt{YMW16} does not model the DM contribution from the Galactic halo, which amounts to $\sim40\text{--}60\,\text{pc}\,\text{cm}^{-3}$ along the line of sight \citep{Yamasaki20}, we still consider these two sources to be of Galactic origin.

\subsection{Timing solutions}

\begin{figure*}
  \includegraphics[width=0.7\linewidth]{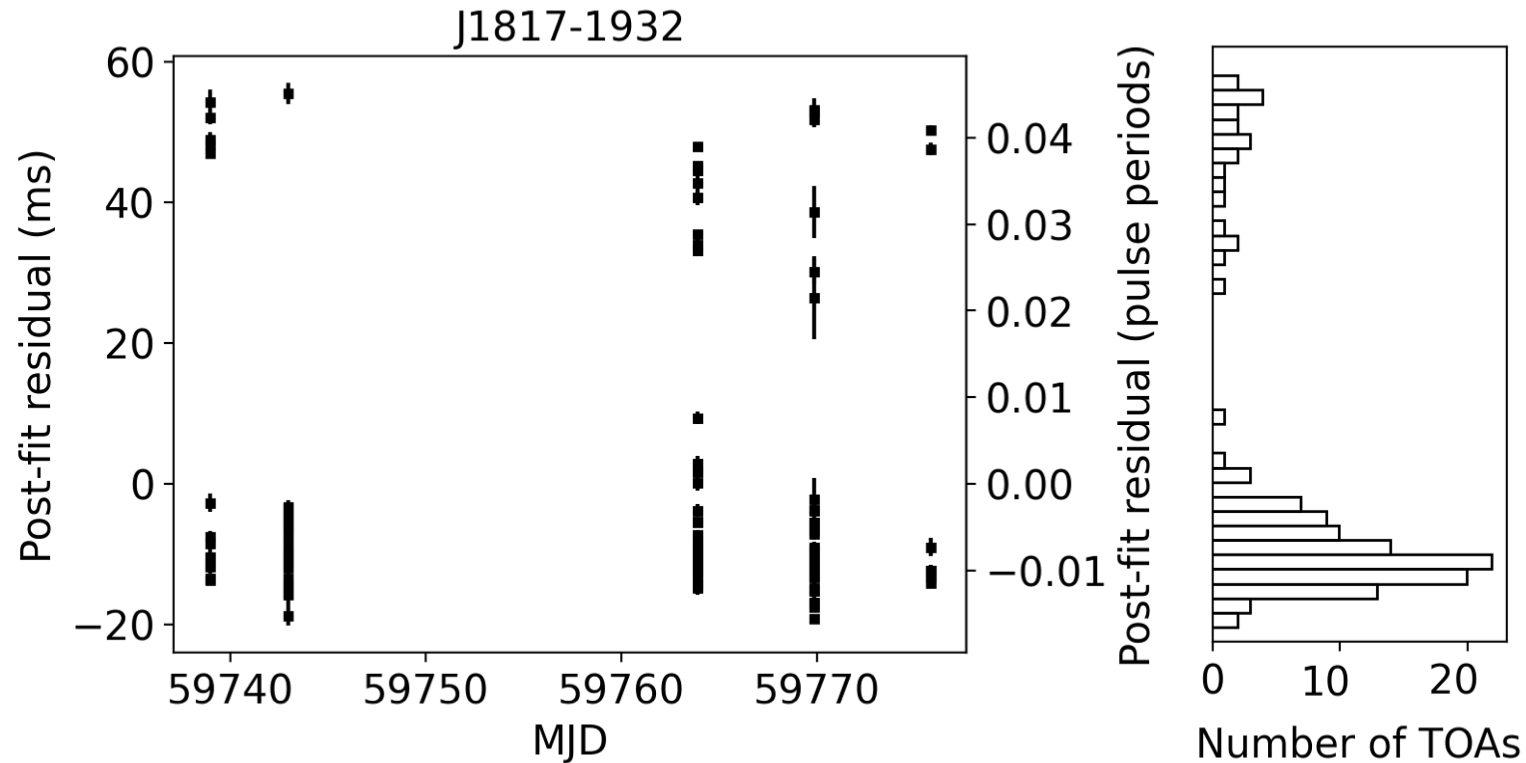}
\caption{Best-fit timing residuals of MTP0063/PSR J1817-1932 (left) and distribution of residuals (right).}
\label{fig:timing}
\end{figure*}

We measured the periods of the MeerTRAP sources with multiple pulses detected closely in time, as listed in Table~\ref{tab:properties}. As MTP0069 and MTP0078 had only two pulses detected in quick succession, we set upper limits on their periods based on the separation between the two pulses. For the sources with pulse detections over a long time span, we used {\sc tempo2} to compare the TOAs against the initial period and imaging localisation, and obtained a phase-connected timing solution for MTP0063. Due to the limited data available for MTP0063, we did not fit any timing parameters other than the period and period derivative. The best-fit timing solution is given in Table~\ref{tab:timing}, and the timing residuals are shown in Figure~\ref{fig:timing}. The period derivative is poorly constrained due to the limited time span of the TOAs. 
The distribution of the timing residuals, as shown in the right panel of Figure~\ref{fig:timing}, suggests two components separated by $\sim50$\,ms in the average profile of MTP0063.
As the radio emission switched between the two components, our methodology of fitting a single Gaussian to the pulse generated the TOA of either the first or the second component. This causes the gap between two sets of TOAs, as can bee seen in Figure~\ref{fig:timing}. This is similar to that observed in some other RRATs discovered by MeerTRAP, e.g. PSRs J1911$-$2020 and J1930$-$1856 \citep{MeerTRAP_james}, and the magnetar-like PSR J1819$-$1458 \citep{Lyne09, Bhattacharyya18}.

\begin{table}
\centering
\caption{Timing solution for MTP0063.}
\label{tab:timing}
\begin{tabular}{ll}
\hline\hline
\multicolumn{2}{c}{Fit and data-set} \\
\hline
PSR name\dotfill & J1817-1932  \\
MeerTRAP name\dotfill & MTP0063  \\
MJD range\dotfill & 59739---59776 \\ 
Number of TOAs\dotfill & 128  \\
rms timing residual (ms)\dotfill & 16.8  \\
Weighted fit\dotfill &  Y   \\ 
Reduced-$\chi^{2}$\dotfill & 780 \\ 
\hline
\multicolumn{2}{c}{Measured Quantities} \\ 
\hline
Right ascension, $\alpha$ (hh:mm:ss)\dotfill & 18:17:12.6(1) \\ 
Declination, $\delta$ (dd:mm:ss)\dotfill & $-$19:32:48(2) \\ 
Spin period, $P$ (s)\dotfill & 1.229084876(7) \\ 
First derivative of $P$, $\dot{P}$ ($\times 10^{-15}$\,ss$^{-1}$)\dotfill & 1(5) \\
Epoch of period determination (MJD)\dotfill & 59769.9 \\ 
Epoch of position determination (MJD)\dotfill & 59769.9 \\ 
Epoch of DM determination (MJD)\dotfill & 59769.9 \\ 
Dispersion measure, DM (cm$^{-3}$pc)\dotfill & 214.5(2) \\ 
\hline
\multicolumn{2}{c}{Assumptions} \\
\hline
Clock correction procedure\dotfill & TT(TAI)  \\
Solar system ephemeris model\dotfill & DE405 \\
Binary model\dotfill & NONE \\
Model version number\dotfill & 5.00 \\ 
\hline
\end{tabular}
\end{table}

\subsection{Fluences}

\begin{figure*}
\subfigure[MTP0055]{
  \includegraphics[width=.45\linewidth]{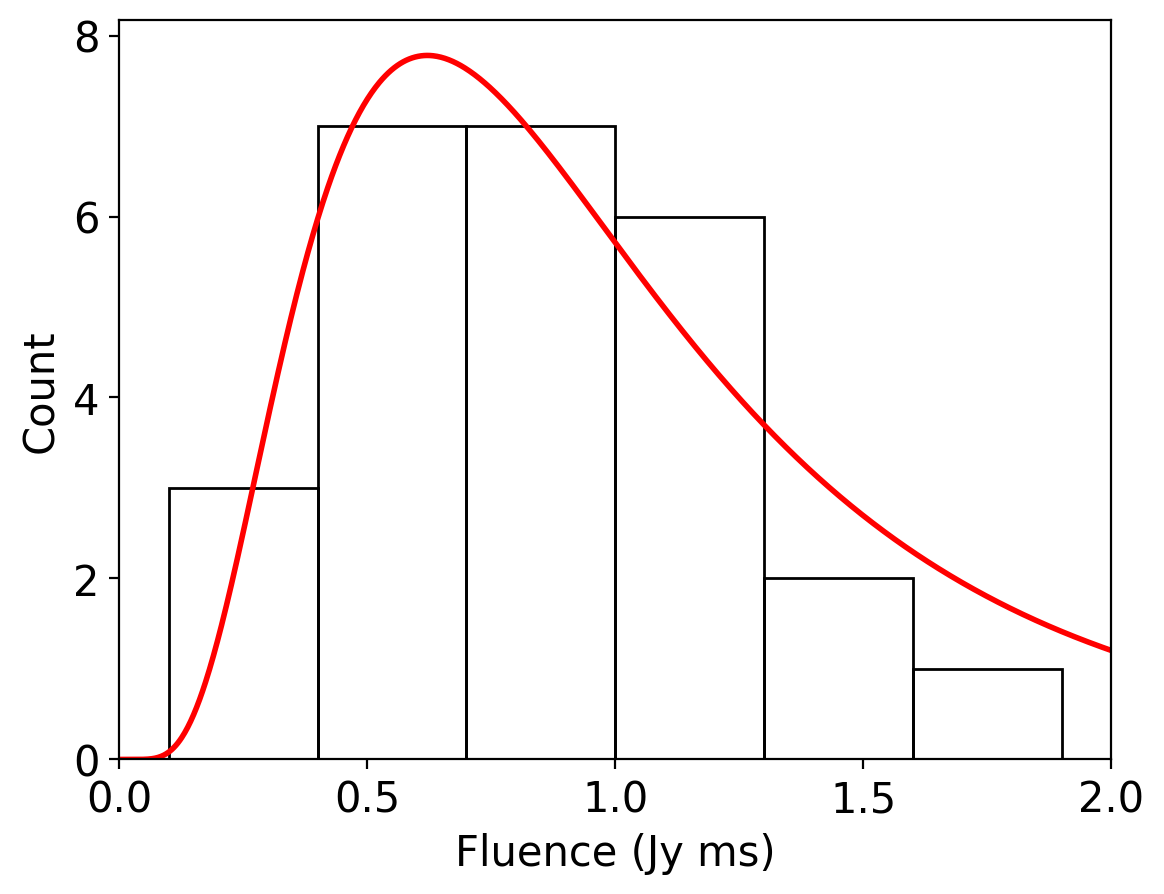}}
\subfigure[MTP0063]{
  \includegraphics[width=.45\linewidth]{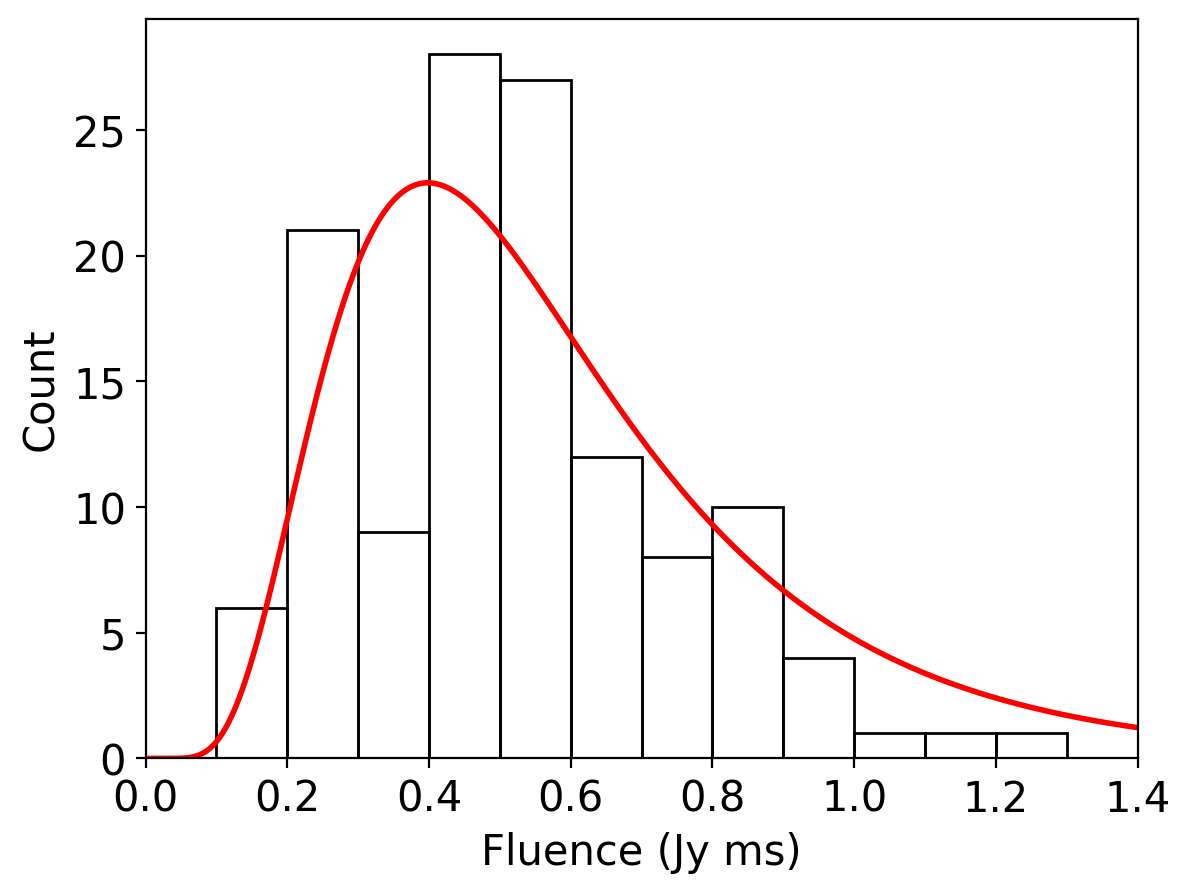}} \\
\subfigure[MTP0067]{
  \includegraphics[width=.45\linewidth]{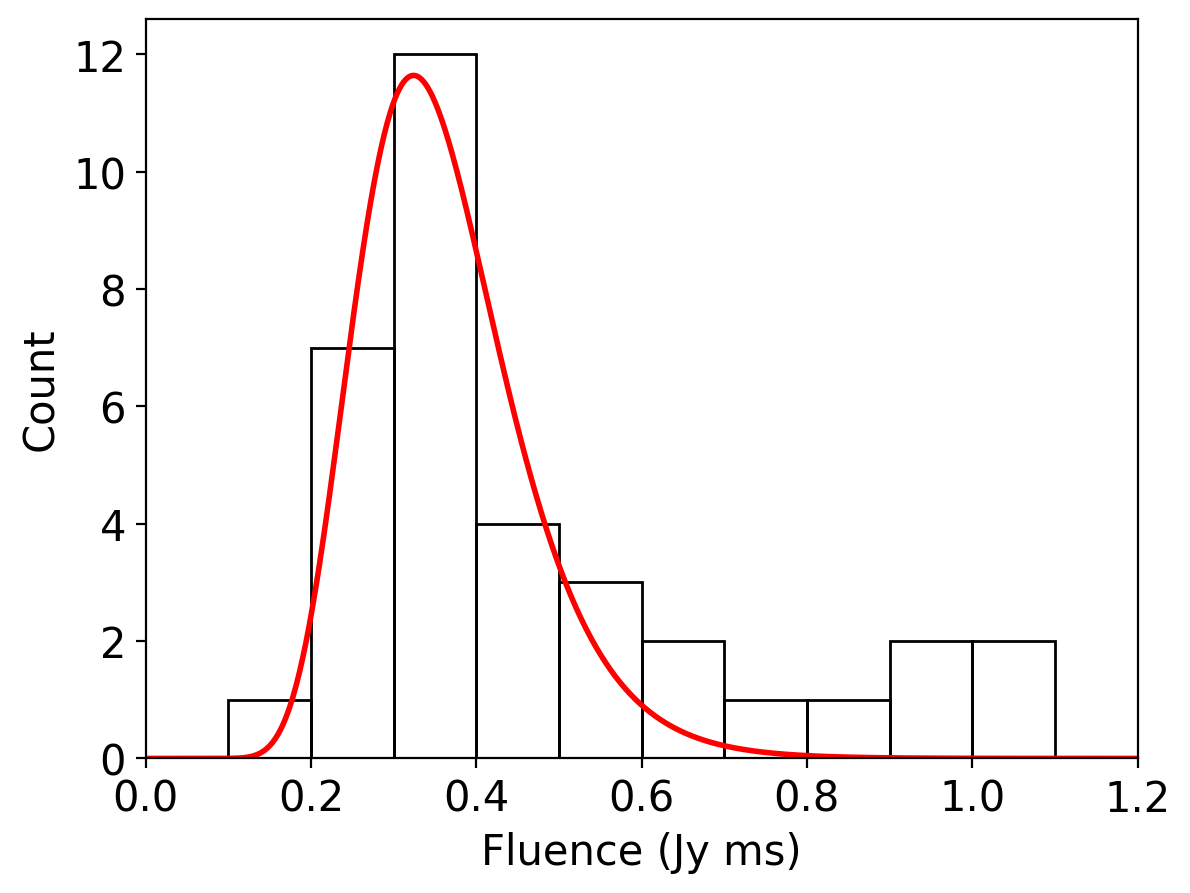}}
\subfigure[MTP0079]{
  \includegraphics[width=.45\linewidth]{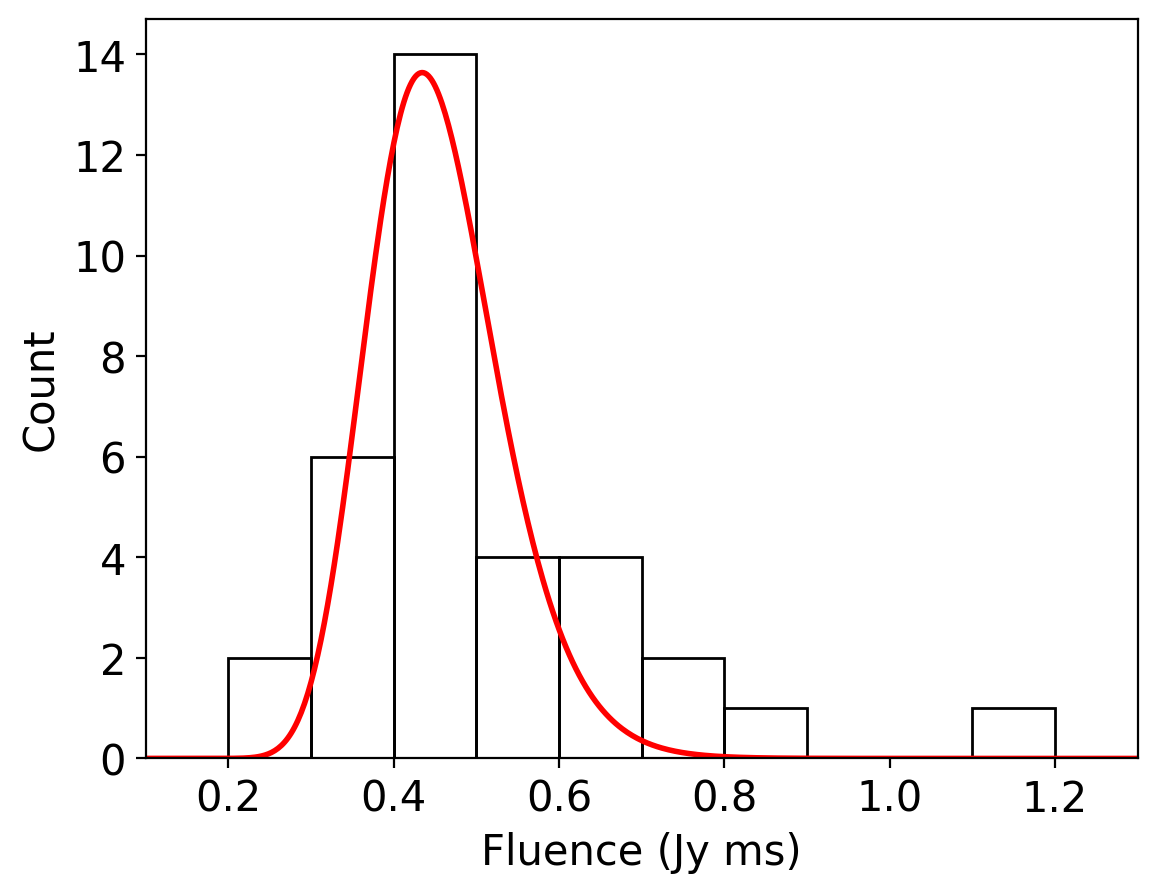}}
\caption{Fluence distributions of individual pulses for the MeerTRAP sources with imaging localisation and more than 20 pulse detections. MTP0055 is detected in both UHF (17 pulses) and L-band (9 pulses), while the other three sources are detected only in the L-band. The red line shows the best-fit lognormal distribution for the observed pulse fluences of each source.}
\label{fig:fluences}
\end{figure*}

We estimated the fluence of individual pulses based on the modified single-pulse radiometer equation (e.g., \citealt{Jankowski23}). This included the calculation of the system equivalent flux density (SEFD) of the MeerKAT array and the array's beam response at the source position. More details can be found in \citet{Tian24b}. The range of fluences measured for individual pulses for each source is given in Table~\ref{tab:properties}. Note that for the sources without imaging localisation, as we do not know their location in the beam, we provide only a lower limit on their fluence. In Figure~\ref{fig:fluences}, we plot the fluence distribution for the sources with $>20$ pulses observed by MeerKAT and with an imaging localisation. MTP0055 has 17 pulses detected at UHF and 9 pulses at L-band, while the other three sources are all at L-band. The fluences of the four sources seem to follow a lognormal like distribution with a peak at $\sim0.3\text{--}0.7$\,Jy\,ms, as shown by the best-fit lognormal distribution in Figure~\ref{fig:fluences}. Previous studies of RRAT pulses also found their fluences to be lognormally distributed \citep{Burke11, Meyers18, Meyers19}. We note that the fluence distributions presented here could be biased due to the MeerTRAP survey performance and completeness limit. 
Given the limited small sample of pulses from the MeerTRAP sources, we cannot obtain the exact fluence distribution here.

\subsection{Polarisation of single pulses}

For the nine sources that triggered voltage buffer dumps we have polarisation information. However, 
many of these sources are very faint and it is hard to get reliable polarisation information and rotation measure (RM) measurements.
Here, we present polarised flux only for those sources with $\text{S/N}>15$ in the detected pulse, i.e. MTP0063, MTP0067, MTP0072 and MTP0074. 
To measure their polarisation, we calibrated the voltage data by applying the primary beam Jones matrix derived from MeerKAT holography experiments \citep{Villiers23}.
A detailed description of the polarisation calibration is given in \citet{Rajwade24}\footnote{The code for performing the polarisation calibration can be found on GitLab: \url{https://gitlab.com/kmrajwade/tbeamformer}}. Note that this method may be inaccurate as there is no measurement of the Jones matrix at the time of the trigger. We therefore added 5 percent uncertainty to the measurement of polarisation fractions to account for any residual leakage.

\begin{figure*}
\subfigure[MTP0063]{
  \includegraphics[width=.35\linewidth]{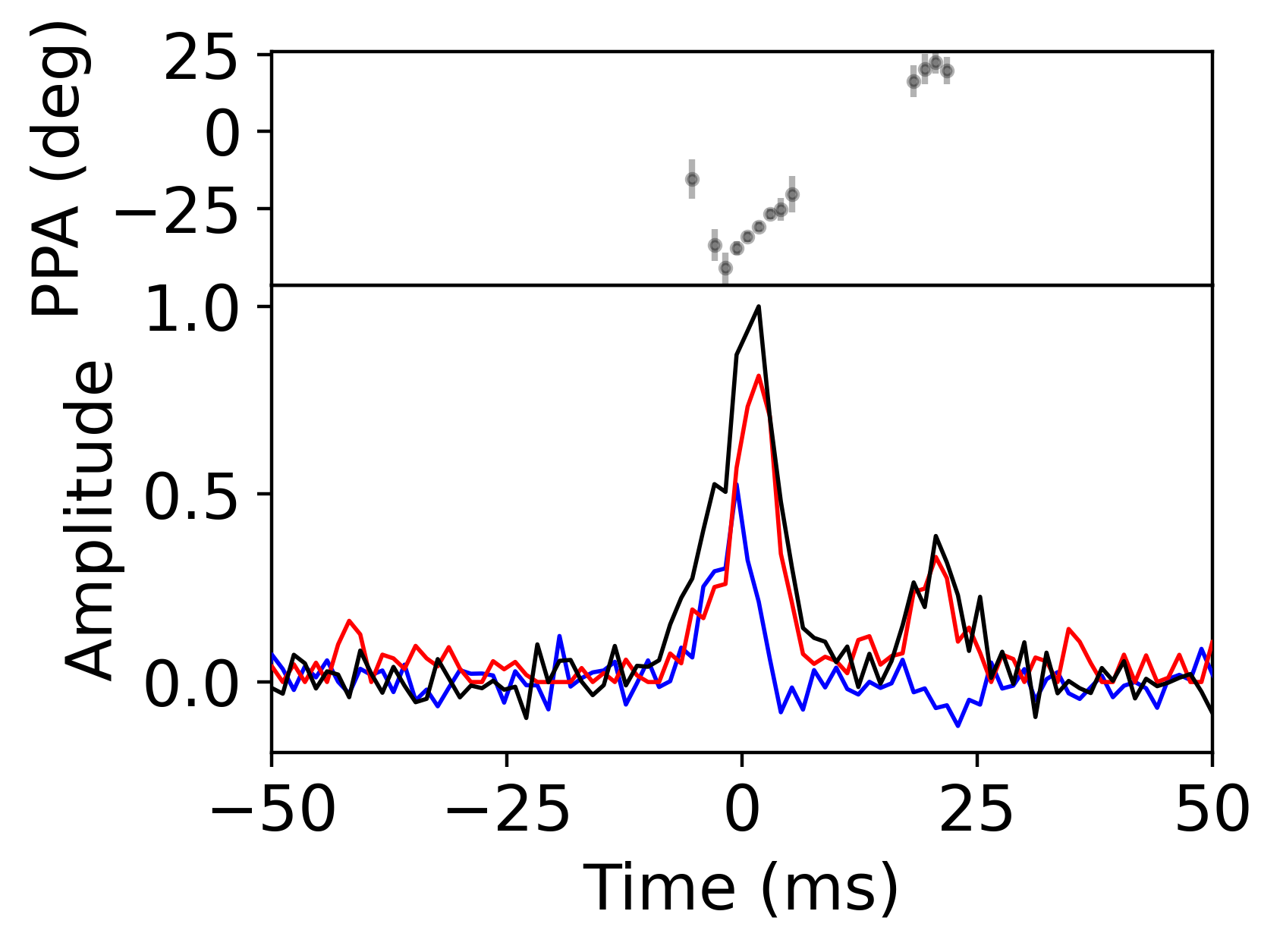}}
\subfigure[MTP0067]{
  \includegraphics[width=.35\linewidth]{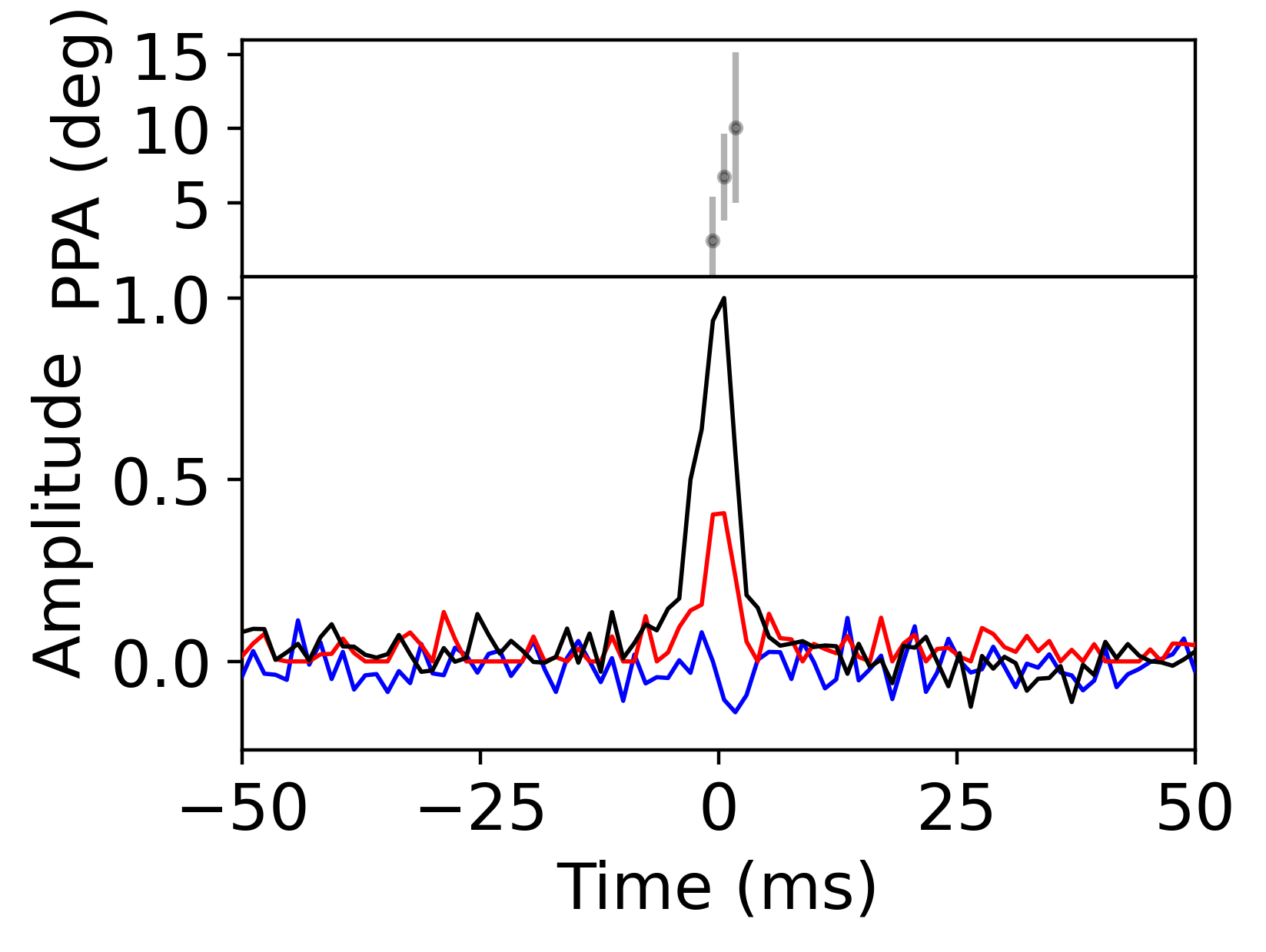}}\\
\subfigure[MTP0072]{
  \includegraphics[width=.35\linewidth]{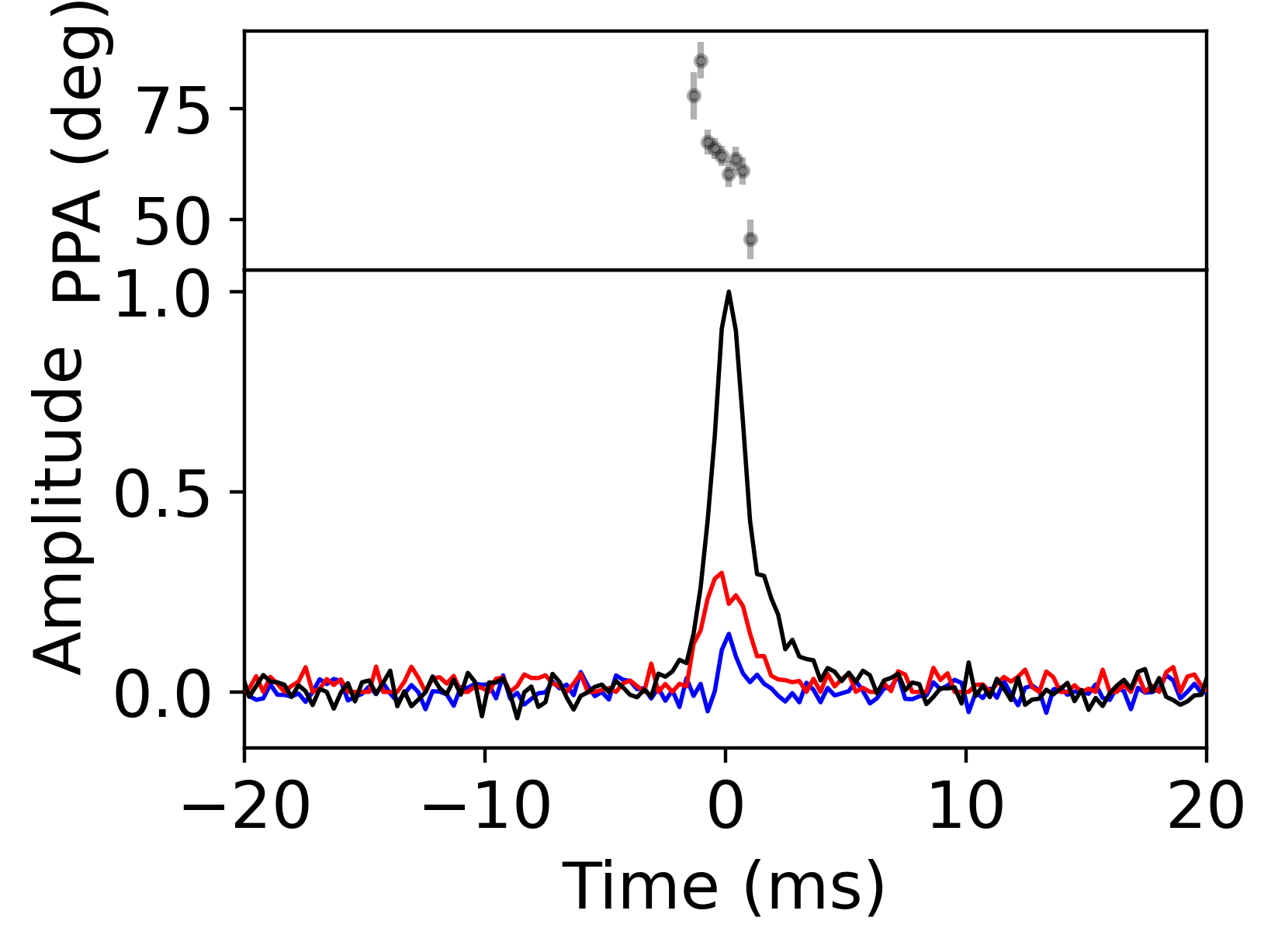}}
\subfigure[MTP0074]{
  \includegraphics[width=.35\linewidth]{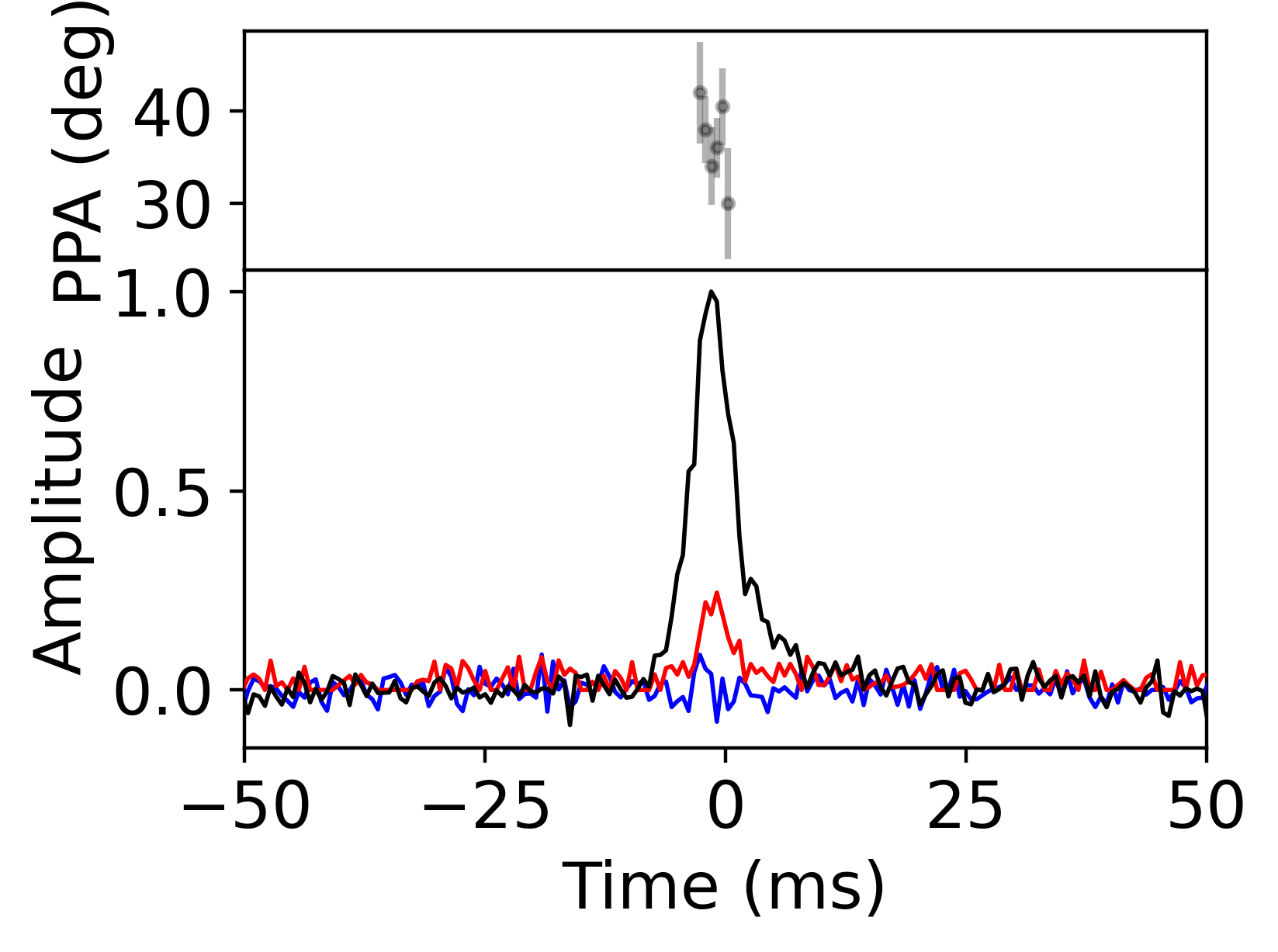}}
\caption{Polarisation profiles of bright pulses created from the transient buffer data. Each panel shows the PPA (top) and the frequency averaged pulse profile (bottom) for total intensity ($I$, black), linear polarisation ($L$, red) and circular polarisation ($V$, blue). The polarisation data are Faraday corrected to the RM values given in Table~\ref{tab:pol}. Vertical error bars indicate the $1\sigma$ uncertainties in the PPA.}
\label{fig:pol}
\end{figure*}

After the calibration, we used the pulsar software {\sc psrsalsa}\footnote{\url{https://github.com/weltevrede/psrsalsa}} \citep{Weltevrede16} to measure the RM and polarisation fraction. We followed the same procedure as described in \citet{Tian24b}. The polarimetric pulse profiles of the four MeerTRAP sources are shown in Figure~\ref{fig:pol}, and the measured RM and linear and circular polarisation fractions are given in Table~\ref{tab:pol}. The pulse of MTP0063 is highly polarised with a linear and circular polarisation fraction of $L/I = 0.79\pm0.13$ and $|V|/I = 0.26\pm0.09$. The other three pulses are less polarised. We see variations of polarisation position angles (PPAs) across these pulse profiles, e.g. sweeping down in MTP0072 and MTP0074 and sweeping up in MTP0063. Such PPA variations are reminiscent of pulsar emission produced in the pulsar magnetosphere \citep[e.g.][]{Han21}.

\begin{table}
\centering
\caption{Polarisation properties of bright pulses for the MeerTRAP sources with transient buffer data.
}
\label{tab:pol}
\begin{tabular}{lccc}
\hline
MTP name & RM & $L/I$ & $|V|/I$ \\
& ($\text{rad}\,\text{m}^{-2}$) & & \\
\hline
MTP0063 & 347(1) & 0.79(13) & 0.26(9)\\
MTP0067 & 76(3) & 0.41(12) & 0.14(10) \\
MTP0072 & 30(2) & 0.43(10) & 0.08(8) \\
MTP0074 & 193(1) & 0.21(7) & 0.06(6) \\
\hline
\end{tabular}
\end{table}

\subsection{UTMOST detection of MTP0068}
\label{subsec:MTP0068_UTMOST_detection}

\begin{figure}
\hspace{-1cm}
\centering
\includegraphics[width=.5\textwidth]{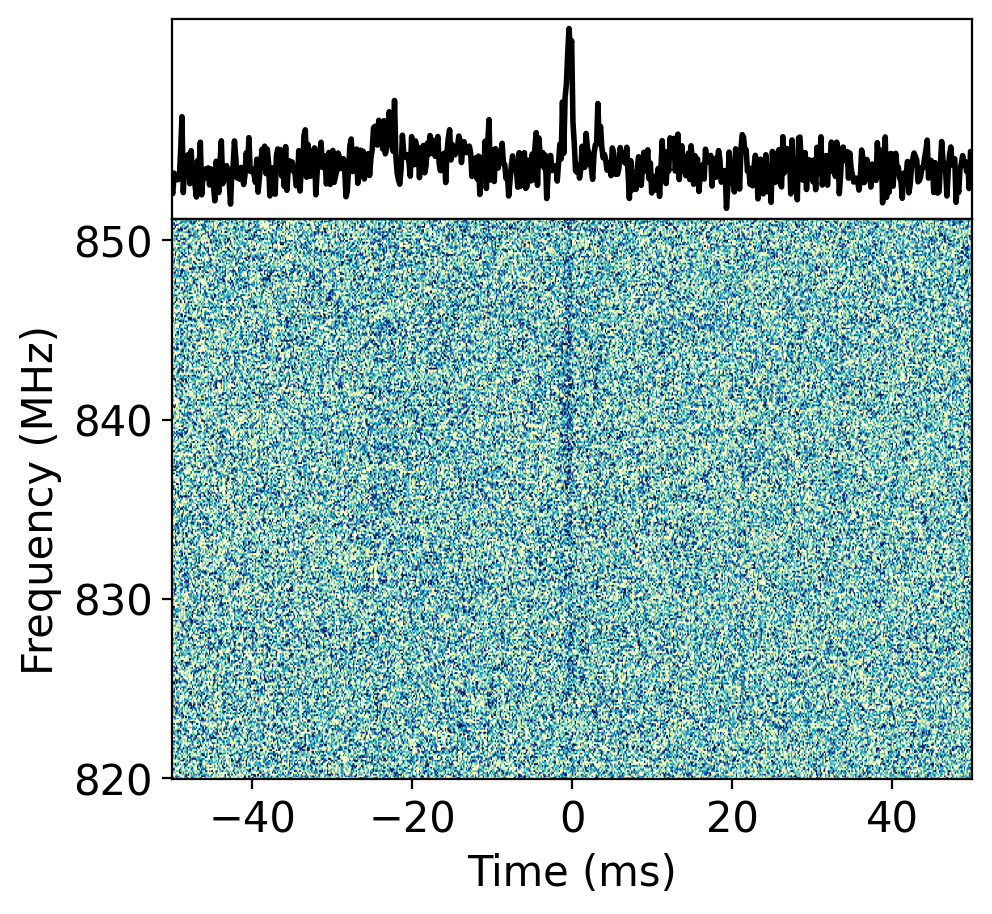}
\caption{The discovery pulse of MTP0068 at UTMOST. The top panel shows the frequency-averaged pulse profile, while the bottom panel shows the dynamic spectrum. The data have been dedispersed to the detection DM of $11.5\,\text{pc}\,\text{cm}^{-3}$.}
\label{fig:molonglo_pulse}
\end{figure}

A source with properties similar to MTP0068 was detected in FRB ``Survey V'', carried out at UTMOST, via its single pulse detection pipeline \citep{Gupta2022}. The source was detected in a 15-minute drift scan observation. The UTMOST survey strategy was to tile its $4\times2$ degree field-of-view with 352 fine stationary phased-array-beams (termed ``fan-beams''). Each fan-beam was independently searched for single pulses in real-time, as the sky drifted through them. 
On 2019 June 18, we detected 3 pulses from an uncatalogued source. Our analysis showed it was moving across the fan-beams at the sky drift rate.
At a detection DM of $11.5\,\text{pc}\,\text{cm}^{-3}$ the dispersion measure of the pulses was well below the DM cutoff for which raw voltage data would be triggered and retained. Subsequently, lower time-resolution ``search-mode'' data (which is retained) were analysed offline. The discovery pulse is shown in Figure~\ref{fig:molonglo_pulse}. The best-fit sky position, consistent for all 3 pulses, was found to be RA=12:43:23.6$\pm45^{''}$ and Dec=$-$05:34:23.2$\pm2^\circ$. Due to the alignment of the UTMOST array along the East-West axis, the precision in localisation is high in the RA direction ($\sim$45 arcsec), but quite poor in the Dec direction ($\sim$2 deg). Using the fan-beam filterbanks of all beams within the field-of-view, we stitched together a tied-array beam filterbank that tracked the best-fit position of the source on the sky.
We then conducted a search for periodicity using the \textsc{riptide} implementation of the FFA \citep{FFA_morello} and 
obtained an $11\sigma$ detection of a 4.86\,s periodicity for the source. 
Given the DM, location and period of this previously uncatalogued source discovered by UTMOST are all consistent with MTP0068, we consider them to be the same source.

\subsection{Microstructure of MTP0068}
Most of the sources reported in this paper have simple morphology with either one or two distinct components. However, MTP0068 (J1243$-$0435) shows interesting pulses with many sub-components. It was discovered by MeerKAT on 2022 September 1 with 5 pulses detected within $\sim30$\,min in the UHF. It has a small DM below the $20\,\text{pc}\,\text{cm}^{-3}$ threshold, and yet was detected (that is why the discovery DM in Table \ref{tab:detections} is $21.49\,\text{pc}\,\text{cm}^{-3}$ compared to the true DM of $12\,\text{pc}\,\text{cm}^{-3}$ given in Table \ref{tab:properties}) due to its brightness and complex sub-structure. As none of the pulses triggered transient buffer data, we utilised the overlap between the detection CBs to constrain the source position. An initial period of 4.867\,s was found using the TOAs of the 5 pulses, and later confirmed using the Effelsberg follow-up observation (see Section~\ref{sec:folding}). This source was also observed by MeerKAT on 2020 September 11 for $\sim1$\,hr and 2022 December 17 for $\sim1$\,hr, but no pulse was detected. This might reflect the clustering of sporadic strong pulses in time.

\begin{figure*}
    \includegraphics[width=\linewidth]{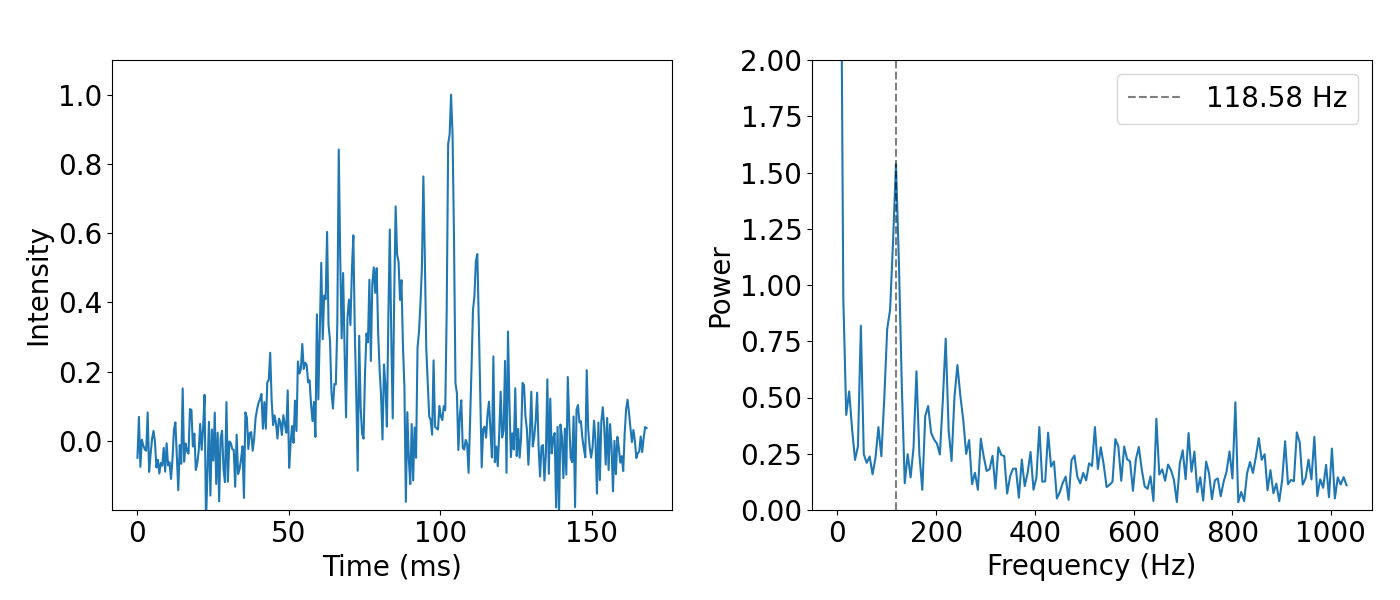}
    \caption{Microstructure in a pulse detected from J1243-0435 on 2022 September 1. The left panel shows the pulse shape, while the right panel shows the power spectrum of the pulse. We find a 3$\sigma$ peak in the power spectrum at a frequency of $118.6^{+9.6}_{-14.4}$\,Hz, corresponding to a microstructure periodicity of $8.4^{+1.2}_{-0.6}$\,ms.}
    \label{fig:micro}
\end{figure*}

The pulses from MTP0068 (J1243$-$0435) show quasi-periodic narrow emission components, known as microstructure \citep{Hankin_review, Mitra_micro}. Microstructure is commonly seen in pulsars, radio magnetars, RRATs, and sometimes even in long-period transients and FRBs \citep{76s_pulsar, Pastor-Marazuela23, microstrucutre_kramer}. 
We only analyzed one bright pulse from J1243$-$0435 (Figure~\ref{fig:micro}, left panel) for microstructure quasi-periodicity; the others were too faint.
We computed the power spectrum of the pulse and identified a peak at 118.6 Hz, as shown in the right panel of Figure~\ref{fig:micro}. 
This peak corresponds to a period of $P_\upmu=8.4^{+1.2}_{-0.6}$~ms, where the uncertainty is determined by the full width at half maximum (FWHM) of the peak. Considering the rotation period of this source $P=4.87$\,s, the microstructure periodicity obtained here is consistent with the empirical relation $P_\upmu\sim 10^{-3}P$ known for radio emitting neutron stars \citep{microstrucutre_kramer}.

In order to assess the significance of this periodicity, we fitted the pulse profile with a multi-component Gaussian model. We allowed the number of components to vary between 7 and 13 and selected the model with the lowest BIC, which in this case corresponded to 10 components. We fitted the arrival time as a function of component number to a linear function, allowing gaps between components, to find the mean inter-component wait-time. Using this mean wait-time, we simulated pulses following a Poissonian wait-time distribution with the same mean spacing as the observed pulses. To prevent simulated components from clustering too closely, we enforced a minimum separation equal to 0.2 times the mean inter-component time, which we refer to as the exclusion parameter, or $\eta$. We then generated $10^5$ simulated sets of times of arrival, and found that 99.881\% of the simulations had larger reduced-$\chi^{2}$ than the tested periodicity. This corresponds to a significance of $3.24\sigma$, which means that the fitted quasi-periodicity is not readily explained by the null hypothesis of Poisson-distributed pulses. More details about the timing technique can be found in \citet{Pastor-Marazuela23}.


\subsection{Persistent emission and folded profiles}\label{sec:folding}
The new sources presented here were found in the real-time MeerTRAP single pulse search. The pipeline detected relatively low number of pulses from most sources, despite extended observation periods when they were in the field of view (see Table~\ref{tab:detections}). However, there could be more regular but weaker emission from these sources. 
In order to find this regular weak signal, we searched the follow-up observations with Effelsberg (see Section \ref{sec:followup}) of five long period (P $\ge$ 4s) MeerTRAP sources. Three of these sources, PSRs J1911$-$2020, J1525$-$2322, and J2218$+$2902, have been published in \citet{MeerTRAP_james} and the other two sources, PSRs J1243$-$0435 (MTP0068) and J1816$-$2419 (MTP0079), are reported here.

\begin{figure*}
\subfigure[J1911–2020]{
  \includegraphics[width=.32\linewidth]{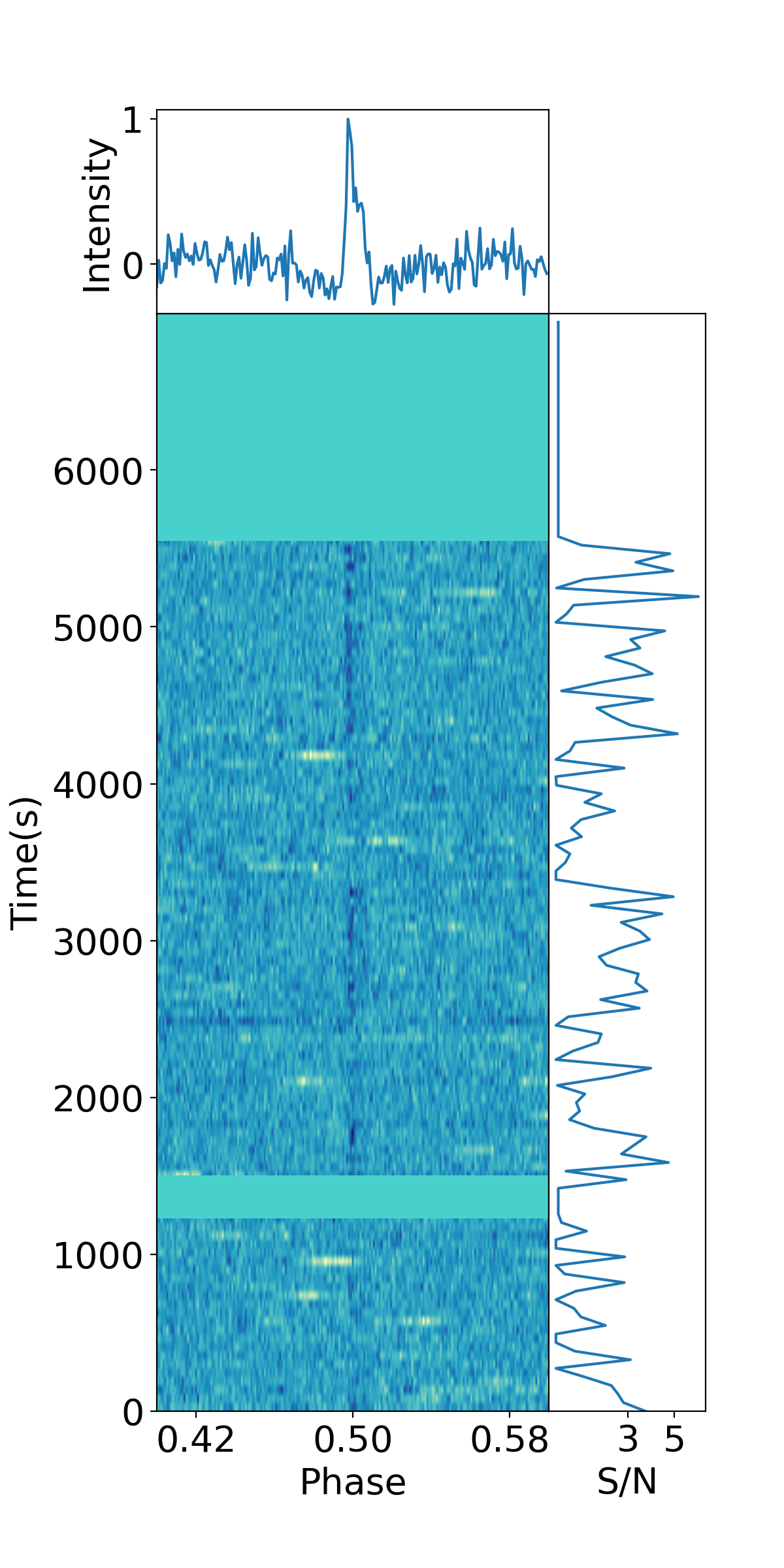}}
\subfigure[J2218+2902]{
  \includegraphics[width=.32\linewidth]{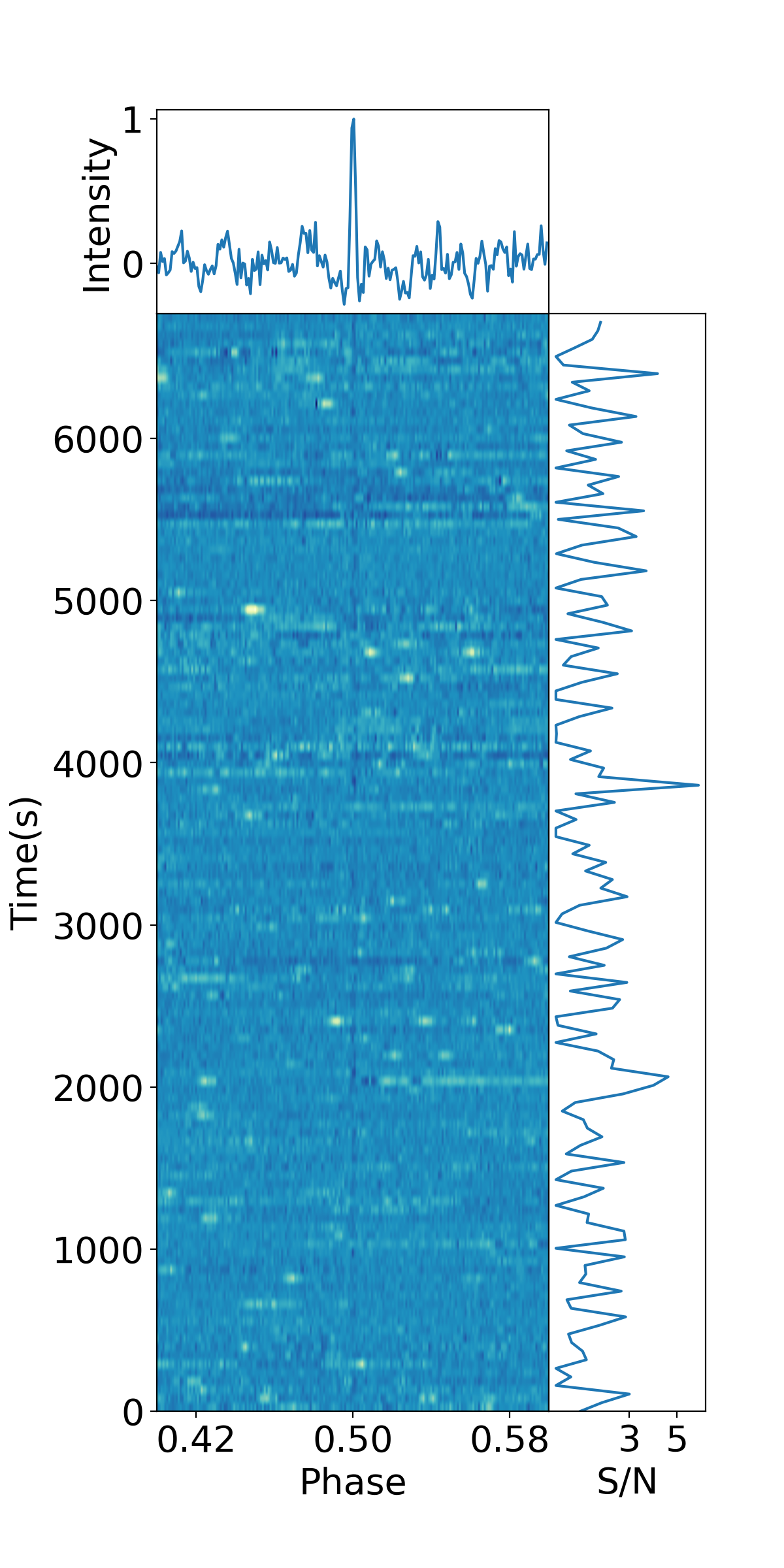}}
\subfigure[J1243-0435]{
  \includegraphics[width=.32\linewidth]{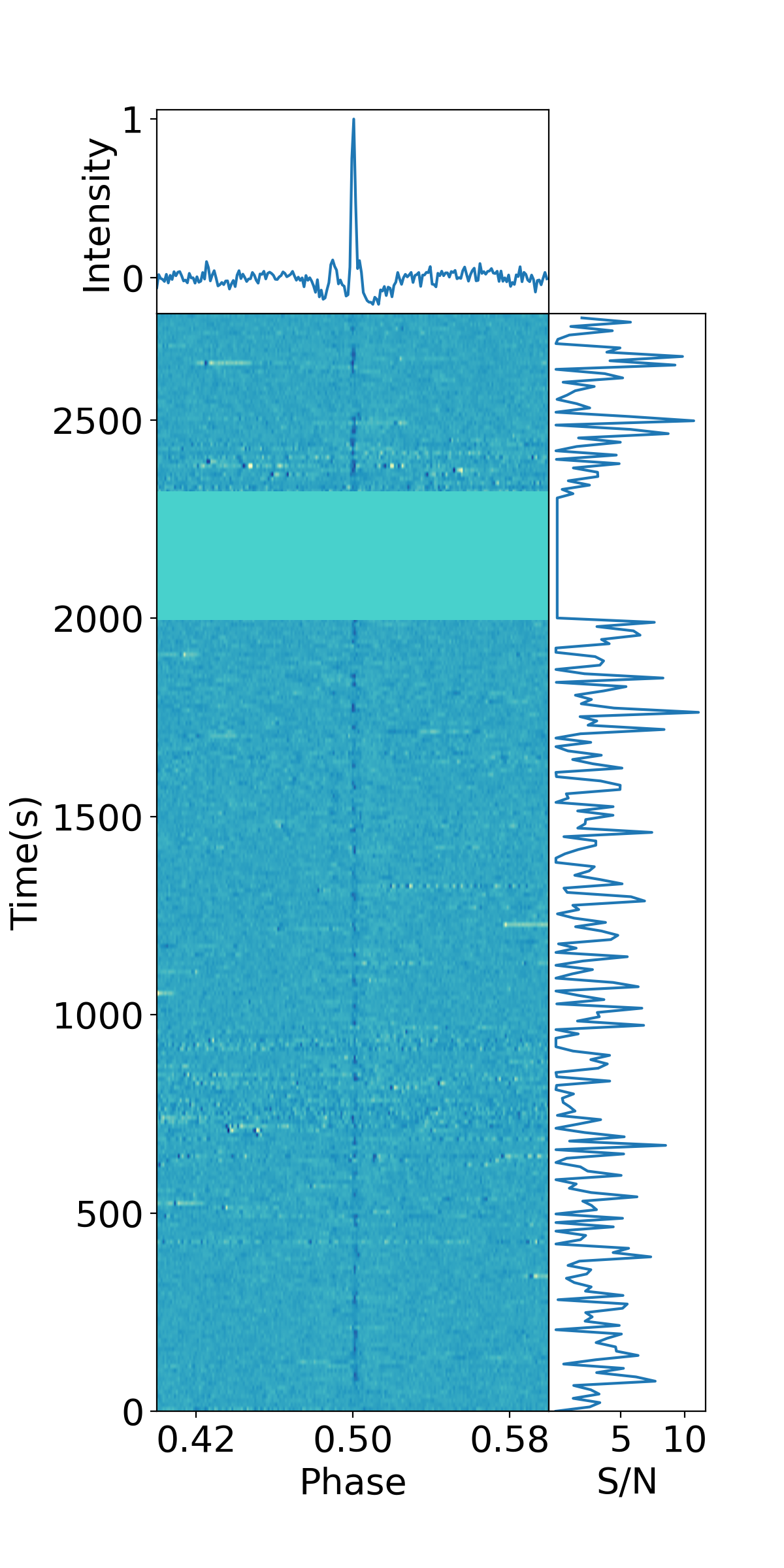}}
\caption{Three MeerTRAP sources detected in the Effelsberg follow-up observations. For each source, the main left panel shows the pulse-stack in the phase-time space, and the top panel shows the averaged pulse profile. The right panel shows the evolution of S/N with time, with the noise calculated from the off-pulse region. Blank horizontal lines in the time-phase plot represent intervals masked due to RFI. The pulsed emission of the three sources is intermittent and shows hints of nulling features.}
\label{fig:subintegration}
\end{figure*}

We found regular periodic signals in the follow-up observations of three sources, PSRs J1911$-$2020, J2218+2902, and J1243$-$0435. Figure \ref{fig:subintegration} shows the phase versus time plot along with the folded profile of these sources, and Table~\ref{tab:duty} lists the duty cycles measured for them.
We can see hints of nulling features in the time-phase plot of these sources. However, due to the limited S/N, we do not measure the nulling fraction here.

\begin{table}
\centering
\caption{Duty cycle and S/N of the pulsed emission measured for the three MeerTRAP sources shown in Figure~\ref{fig:subintegration}.
}
\label{tab:duty}
\begin{tabular}{lcc}
\hline
PSR name & Duty cycle & S/N \\
\hline
J1911$-$2020 & $(0.84\pm0.10)\%$ & 20.9 \\
J2218+2902 & $(0.29\pm0.10)\%$ & 27.4 \\
J1243$-$0435 & $(0.21\pm0.02)\%$ & 81.4 \\
\hline
\end{tabular}
\end{table}

\section{Discussion}\label{sec:disc}

\subsection{Interesting sources}


Many of the MeerTRAP sources have been detected only a few times despite the large amount of observing time in the IB mode at MeerKAT, as can be seen in Table~\ref{tab:detections}. This could be due to an intrinsically low activity of these sources or their pulses being too faint to be detected by the IB mode. More sensitive follow-ups are needed to confirm whether these sources are genuinely inactive or their pulses are too faint.

A few of the Galactic sources presented here exhibit unusual properties. PSR J2218+2902 was initially reported in \citet{MeerTRAP_james}. 
There are only a few slow pulsars discovered so far, including PSR J0250+5854 ($\text{P}=23.5$\,s; \citealt{23s_pulsar}), J0901-4046 ($\text{P}=76$\,s; \citealt{76s_pulsar}) and J0311+1402 ($\text{P}=41$\,s; \citealt{WangY25}). Searching for more of them is important as they can be used to constrain the pulsar emission mechanism and provide new insights into the neutron star evolution path. Our follow-up of PSR J2218+2902 with Effelsberg found faint periodic emission with a period of 17.49\,s (see Figure~\ref{fig:subintegration}).
Therefore, we confirm PSR J2218+2902 to be the fourth slowest in the radio pulsar population.

Three out of the five sources we have followed up with Effelsberg turn out to be also detectable as apparently normal pulsars.
These three sources are all located below the Galactic plane with $|b|>10^\circ$ and thus not covered by Galactic plane pulsar surveys (e.g.~\citealt{Manchester01, Keane18, Hessels08, Cordes06}). PSRs J1911$-$2020 and J1243$-$0435 are within the region of the HTRU-South survey, which targets the entire sky south of declination of $+10^\circ$ \citep{Keith10}. However, these two sources were not discovered in this survey probably because of the red noise in the data, which reduced the sensitivity to pulsars with a spin period above 500\,ms or the flux density of the two sources being below the survey sensitivity.
It is possible that some other Galactic sources discovered by MeerTRAP are pulsars with very sparse bright pulses as well. Continuous follow-up of more MeerTRAP sources would reveal more pulsars in the apparent RRAT population and enable studies of their pulse energy distributions and further elucidate the nature of this population of radio emitting neutron stars and how they fit in with the more steady emitting pulsars.

We note the extended source to the North of MTP0079 is the supernova remnant (SNR) G7.7$-$3.7 \citep[see e.g.][]{Cotton24}. We briefly investigated whether MTP0079 and G7.7$-$3.7 could have been produced by the same supernova event. This scenario would mean MTP0079 is much younger than other neutron stars discovered by MeerTRAP. The lifetime of radio detected SNRs is typically $\lesssim80$\,kyr \citep{Frail94, Vink20}, and G7.7$-$3.7 has been suggested to be a possible remnant of the supernova of AD386 \citep{Zhou18}. The offset of MTP0079 from the geometric center of G7.7$-$3.7 is $\approx18.2$\,arcmin. Assuming the \texttt{NE2001} distance of 6.1\,kpc for MTP0079, its birth velocity would be unprecedentedly large ($\sim2\times10^{4}\,\text{km}\,\text{s}^{-1}$) if it was associated with the supernova of AD386. 
however the derived velocity may be wrong given that G7.7$-$3.7 has not been extensively studied and its age is very uncertain.
However, this region near the Galactic plane is a highly populated area of the sky. Therefore, measurements of the proper motion of MTP0079 would be needed to firmly establish an association.

\begin{figure*}
\hspace{-1cm}
\centering
\includegraphics[width=.7\textwidth]{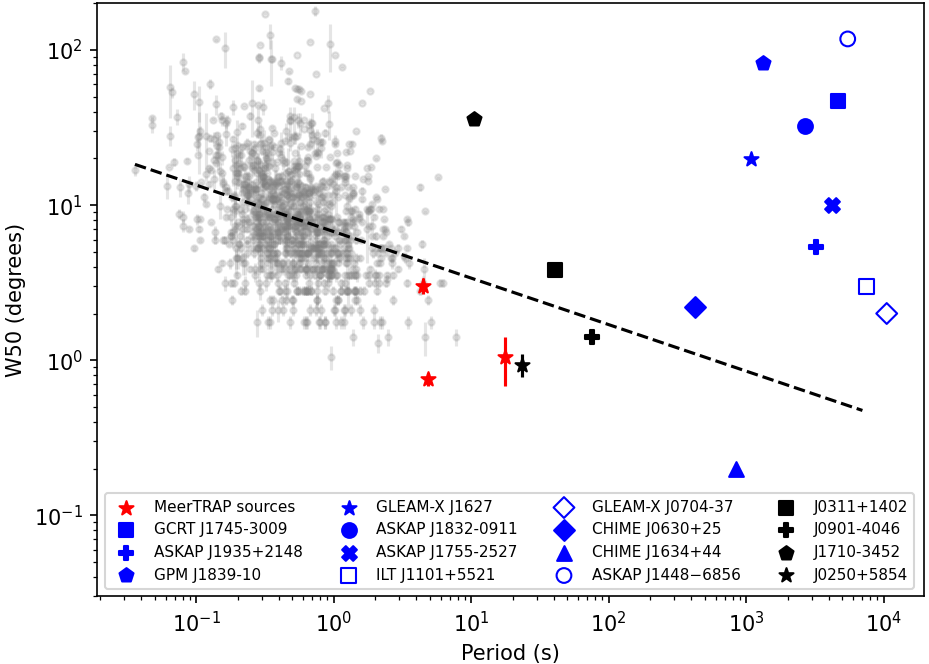}
\caption{Duty cycle vs. pulsar period. The gray dots represent pulsars from the Thousand-Pulsar-Array programme on MeerKAT \citep{Posselt21}, the black points represent the three long-period pulsars \citep{23s_pulsar, 76s_pulsar, Surnis23, WangY25}, the blue points represent the ULP transients \citep{Hyman05, Caleb24, Hurley22, 18m_source, Wang24, McSweeney25b, Ruiter25, Hurley24, Dong25b, Dong25, Anumarlapudi25}, and the red points represent the three MeerTRAP sources with persistent emission observed by Effelsberg (see Section~\ref{sec:folding}). We indicate the ULPs with optical counterparts in empty markers. The dashed line shows the empirical relation $W\propto P^{-0.3}$ between the observed profile width and pulsar spin period \citep{Johnston19}.}
\label{fig:duty_cycle}
\end{figure*}

\subsection{Future prospects}
The discoveries reported here and in \citet{Bezuidenhout22} and \citet{MeerTRAP_james} demonstrate the capability of the single-pulse search program with the sensitive telescope MeerKAT. Such a campaign can not only increase the population of FRBs but also significantly contribute to the population of Galactic transients. 
While Fast Fourier Transform (FFT) based periodicity searches have discovered the majority of known pulsars, they are known to be less sensitive towards pulsars with long periods and/or small duty cycles \citep{heerden_2016, FFA_morello}.
An alternative periodicity search method, called the Fast Folding Algorithm (FFA) search \citep{staelin}, provides better sensitivity for pulsars with long periods and small duty cycles \citep{FFA_morello, singh_FFA}. The FFA search is also believed to be more sensitive towards highly nulling pulsars \citep{GHRSS_VI, grover_FFA}. However, these periodicity search methods may still miss weak pulsars with occasional bright pulses, as demonstrated by \citet{Zhou23}. These pulsars are more likely to be detected by single pulse searches like used by MeerTRAP and other projects searching for FRBs. 
In addition, we have discovered some sources with long periods and small duty cycles.
Figure~\ref{fig:duty_cycle} shows the duty cycle of these sources, in comparison with radio pulsars \citep{Posselt21}, long-period pulsars \citep{23s_pulsar, Surnis23, WangY25} and ULP transients \citep{Hyman05, 76s_pulsar, Caleb24, Hurley22, 18m_source, Wang24, McSweeney25b, Ruiter25, Hurley24, Dong25b, Dong25, Anumarlapudi25}. As can be seen, the duty cycles of the MeerTRAP sources are very low, which might imply a narrow pulsar beam, however this depends on the viewing geometry. While the long-period pulsars follow the general trend for the duty cycle with period \citep{Maciesiak12, Skrzypczak18, Posselt21}, the ULP transients look like they are outliers and some of them may not be neutron stars, but white dwarf binaries, e.g. GLEAM-X J0704$-$37 and ILT J1101+5521 \citep{Hurley24, Ruiter25}. With MeerTRAP we expect to discover more long-period pulsars that are located in the gap between pulsars and long-period transients, which would further reveal their relationships.

The results from the follow-up observations presented here motivate us to follow up more Galactic transients discovered by MeerTRAP with other telescopes. These follow-up observations will help us obtain a coherent timing solution, allowing us to determine their spin down rate, magnetic field strength and characteristic age, potentially placing these discoveries in the under-populated regions of the $P-\dot{P}$ diagram (e.g. the magnetar regime or the pulsar death valley). As can be seen in this work, the long continuous observations can sometimes reveal weak regular emission from the apparently transient sources. Follow-up with sensitive telescopes may also allow us to study their emission properties in greater detail.

\section{Conclusions}\label{sec:concl}

In this paper, we have presented 30 new Galactic sources the MeerTRAP project discovered in real-time via a single pulse search. This work, along with that published by \citet{Bezuidenhout22} and \citet{MeerTRAP_james}, have demonstrated the efficiency of MeerTRAP in finding Galactic transient sources. Most of these sources are likely to be RRATs given their low pulse rates. We localised nine of the MeerTRAP sources to arcsecond precision using the voltage buffer data, and obtained a phase-coherent timing solution for MTP0063/PSR J1817-1932. 
We estimated the fluence distribution of single pulses for four of the MeerTRAP sources. The fluences range between $\sim0.1\text{--}2$\,Jy\,ms with a peak at $\sim0.3\text{--}0.7$\,Jy\,ms, and might follow a lognormal distribution. We also measured polarisation properties of single pulses for four of the MeerTRAP sources, and found them to be linearly polarised with a polarisation fraction up to 79 per cent. We also investigated the microstructure observed in PSR J1243-0435. A period of $8.4^{+1.2}_{-0.6}$~ms was identified at $3\sigma$ significance in the brightest pulse from PSR J1243-0435. This microstructure periodicity and the spin period of the source follow the universal relation previously seen in radio emitting neutron stars \cite{microstrucutre_kramer}. In addition, we found regular faint emission from three sources using the Effelsberg follow-up observations. This confirms their long rotation period, including PSR J2218+2902 with a period of 17.5\,s. Considering the ongoing transient search with MeerTRAP, we expect to discover more RRATs and pulsars, especially those with long rotation periods.






\section*{Acknowledgements}
The MeerKAT telescope is operated by the South African Radio Astronomy Observatory (SARAO), which is a facility of the National Research Foundation, itself an agency of the Department of Science and Innovation. All the authors thank the MeerKAT LSP teams, including MALS, MeerTime, MHONGOOSE and ThunderKAT, for allowing commensal observing and the staff at SARAO for scheduling MeerKAT observations. MeerTRAP observations use the FBFUSE and TUSE computing clusters for data acquisition and storage. These instruments were designed, funded and installed by the Max-Planck Institut f{\"u}r Radioastronomie (MPIfR) and the Max-Planck-Gesellschaft. 
This publication includes data from observations with the 100-m telescope of the MPIfR at Effelsberg. The Effelsberg UBB receiver and EDD system are developed and maintained by the electronics division at the MPIfR. The authors are thankful of Dr. Alex Kraus for scheduling the observations.
MeerTRAP acknowledges funding from the European Research Council (ERC) under the European Union’s Horizon 2020 research and innovation programme (grant agreement no. 694745). JT and BWS acknowledge funding from an STFC Consolidated grant. JDT acknowledges funding from the United Kingdom's Research and Innovation Science and Technology Facilities Council (STFC) Doctoral Training Partnership, project code 2659479. M.C. acknowledges support of an Australian Research Council Discovery Early Career Research Award (project number DE220100819) funded by the Australian Government.
IPM further acknowledges funding from an NWO Rubicon Fellowship, project number 019.221EN.019.
For the purpose of open access, the author has applied a Creative Commons Attribution (CC BY) licence to any Author Accepted Manuscript version arising.
This research used version 2.2.0 of the ATNF Pulsar Catalogue.
This research has made use of the SIMBAD data base, operated at CDS, Strasbourg, France \citep{Wenger2000}. This research has made use of NASA’s Astrophysics Data System Bibliographic Services.
\section*{Data Availability}
The data underlying this article will be shared on reasonable request to the corresponding authors.


\bibliographystyle{mnras}
\bibliography{references,bibfile} 

\begin{thebibliography}{}
\makeatletter
\relax
\def\mn@urlcharsother{\let\do\@makeother \do\$\do\&\do\#\do\^\do\_\do\%\do\~}
\def\mn@doi{\begingroup\mn@urlcharsother \@ifnextchar [ {\mn@doi@} {\mn@doi@[]}}
\def\mn@doi@[#1]#2{\def\@tempa{#1}\ifx\@tempa\@empty \href {http://dx.doi.org/#2} {doi:#2}\else \href {http://dx.doi.org/#2} {#1}\fi \endgroup}
\def\mn@eprint#1#2{\mn@eprint@#1:#2::\@nil}
\def\mn@eprint@arXiv#1{\href {http://arxiv.org/abs/#1} {{\tt arXiv:#1}}}
\def\mn@eprint@dblp#1{\href {http://dblp.uni-trier.de/rec/bibtex/#1.xml} {dblp:#1}}
\def\mn@eprint@#1:#2:#3:#4\@nil{\def\@tempa {#1}\def\@tempb {#2}\def\@tempc {#3}\ifx \@tempc \@empty \let \@tempc \@tempb \let \@tempb \@tempa \fi \ifx \@tempb \@empty \def\@tempb {arXiv}\fi \@ifundefined {mn@eprint@\@tempb}{\@tempb:\@tempc}{\expandafter \expandafter \csname mn@eprint@\@tempb\endcsname \expandafter{\@tempc}}}

\bibitem[\protect\citeauthoryear{{Anumarlapudi} et~al.,}{{Anumarlapudi} et~al.}{2025}]{Anumarlapudi25}
{Anumarlapudi} A.,  et~al., 2025, \mn@doi [\mnras] {10.1093/mnras/staf1227}, \href {https://ui.adsabs.harvard.edu/abs/2025MNRAS.542.1208A} {542, 1208}

\bibitem[\protect\citeauthoryear{{Archibald} et~al.,}{{Archibald} et~al.}{2017}]{Archibald17}
{Archibald} R.~F.,  et~al., 2017, \mn@doi [\apjl] {10.3847/2041-8213/aa9371}, \href {https://ui.adsabs.harvard.edu/abs/2017ApJ...849L..20A} {849, L20}

\bibitem[\protect\citeauthoryear{{Barr} et~al.,}{{Barr} et~al.}{2013}]{Barr13}
{Barr} E.~D.,  et~al., 2013, \mn@doi [\mnras] {10.1093/mnras/stt1440}, \href {https://ui.adsabs.harvard.edu/abs/2013MNRAS.435.2234B} {435, 2234}

\bibitem[\protect\citeauthoryear{{Beskin}, {Chernov}, {Gwinn}  \& {Tchekhovskoy}}{{Beskin} et~al.}{2015}]{Pulsar_review_beskin}
{Beskin} V.~S.,  {Chernov} S.~V.,  {Gwinn} C.~R.,   {Tchekhovskoy} A.~A.,  2015, \mn@doi [\ssr] {10.1007/s11214-015-0173-8}, \href {https://ui.adsabs.harvard.edu/abs/2015SSRv..191..207B} {191, 207}

\bibitem[\protect\citeauthoryear{{Bezuidenhout} et~al.,}{{Bezuidenhout} et~al.}{2022}]{Bezuidenhout22}
{Bezuidenhout} M.~C.,  et~al., 2022, \mn@doi [\mnras] {10.1093/mnras/stac579}, \href {https://ui.adsabs.harvard.edu/abs/2022MNRAS.512.1483B} {512, 1483}

\bibitem[\protect\citeauthoryear{{Bezuidenhout} et~al.,}{{Bezuidenhout} et~al.}{2023}]{Bezuidenhout23}
{Bezuidenhout} M.~C.,  et~al., 2023, \mn@doi [RAS Techniques and Instruments] {10.1093/rasti/rzad007}, \href {https://ui.adsabs.harvard.edu/abs/2023RASTI...2..114B} {2, 114}

\bibitem[\protect\citeauthoryear{{Bhat} et~al.,}{{Bhat} et~al.}{2023a}]{Bhat23b}
{Bhat} N.~D.~R.,  et~al., 2023a, \mn@doi [\pasa] {10.1017/pasa.2023.18}, \href {https://ui.adsabs.harvard.edu/abs/2023PASA...40...20B} {40, e020}

\bibitem[\protect\citeauthoryear{{Bhat} et~al.,}{{Bhat} et~al.}{2023b}]{Bhat23a}
{Bhat} N.~D.~R.,  et~al., 2023b, \mn@doi [\pasa] {10.1017/pasa.2023.17}, \href {https://ui.adsabs.harvard.edu/abs/2023PASA...40...21B} {40, e021}

\bibitem[\protect\citeauthoryear{{Bhattacharyya} et~al.,}{{Bhattacharyya} et~al.}{2018}]{Bhattacharyya18}
{Bhattacharyya} B.,  et~al., 2018, \mn@doi [\mnras] {10.1093/mnras/sty923}, \href {https://ui.adsabs.harvard.edu/abs/2018MNRAS.477.4090B} {477, 4090}

\bibitem[\protect\citeauthoryear{{Bhattacharyya} et~al.,}{{Bhattacharyya} et~al.}{2019}]{Bhattacharyya19}
{Bhattacharyya} B.,  et~al., 2019, \mn@doi [\apj] {10.3847/1538-4357/ab2bf3}, \href {https://ui.adsabs.harvard.edu/abs/2019ApJ...881...59B} {881, 59}

\bibitem[\protect\citeauthoryear{{Bochenek}, {Ravi}, {Belov}, {Hallinan}, {Kocz}, {Kulkarni}  \& {McKenna}}{{Bochenek} et~al.}{2020}]{Bochenek20}
{Bochenek} C.~D.,  {Ravi} V.,  {Belov} K.~V.,  {Hallinan} G.,  {Kocz} J.,  {Kulkarni} S.~R.,   {McKenna} D.~L.,  2020, \mn@doi [\nat] {10.1038/s41586-020-2872-x}, \href {https://ui.adsabs.harvard.edu/abs/2020Natur.587...59B} {587, 59}

\bibitem[\protect\citeauthoryear{{Burke-Spolaor} et~al.,}{{Burke-Spolaor} et~al.}{2011}]{Burke11}
{Burke-Spolaor} S.,  et~al., 2011, \mn@doi [\mnras] {10.1111/j.1365-2966.2011.18521.x}, \href {https://ui.adsabs.harvard.edu/abs/2011MNRAS.416.2465B} {416, 2465}

\bibitem[\protect\citeauthoryear{{CHIME Collaboration} et~al.,}{{CHIME Collaboration} et~al.}{2022}]{CHIME22}
{CHIME Collaboration} et~al., 2022, \mn@doi [\apjs] {10.3847/1538-4365/ac6fd9}, \href {https://ui.adsabs.harvard.edu/abs/2022ApJS..261...29C} {261, 29}

\bibitem[\protect\citeauthoryear{{CHIME/FRB Collaboration} et~al.,}{{CHIME/FRB Collaboration} et~al.}{2018}]{CHIME18}
{CHIME/FRB Collaboration} et~al., 2018, \mn@doi [\apj] {10.3847/1538-4357/aad188}, \href {https://ui.adsabs.harvard.edu/abs/2018ApJ...863...48C} {863, 48}

\bibitem[\protect\citeauthoryear{{CHIME/FRB Collaboration} et~al.,}{{CHIME/FRB Collaboration} et~al.}{2019}]{CHIME19b}
{CHIME/FRB Collaboration} et~al., 2019, \mn@doi [\apjl] {10.3847/2041-8213/ab4a80}, \href {https://ui.adsabs.harvard.edu/abs/2019ApJ...885L..24C} {885, L24}

\bibitem[\protect\citeauthoryear{{CHIME/FRB Collaboration} et~al.,}{{CHIME/FRB Collaboration} et~al.}{2020}]{CHIME20}
{CHIME/FRB Collaboration} et~al., 2020, \mn@doi [\nat] {10.1038/s41586-020-2863-y}, \href {https://ui.adsabs.harvard.edu/abs/2020Natur.587...54C} {587, 54}

\bibitem[\protect\citeauthoryear{{Caleb} et~al.,}{{Caleb} et~al.}{2019}]{RRAT_polarization}
{Caleb} M.,  et~al., 2019, \mn@doi [\mnras] {10.1093/mnras/stz1352}, \href {https://ui.adsabs.harvard.edu/abs/2019MNRAS.487.1191C} {487, 1191}

\bibitem[\protect\citeauthoryear{{Caleb} et~al.,}{{Caleb} et~al.}{2022}]{76s_pulsar}
{Caleb} M.,  et~al., 2022, \mn@doi [Nature Astronomy] {10.1038/s41550-022-01688-x}, \href {https://ui.adsabs.harvard.edu/abs/2022NatAs...6..828C} {6, 828}

\bibitem[\protect\citeauthoryear{{Caleb} et~al.,}{{Caleb} et~al.}{2023}]{Caleb23}
{Caleb} M.,  et~al., 2023, \mn@doi [\mnras] {10.1093/mnras/stad1839}, \href {https://ui.adsabs.harvard.edu/abs/2023MNRAS.524.2064C} {524, 2064}

\bibitem[\protect\citeauthoryear{{Caleb} et~al.,}{{Caleb} et~al.}{2024}]{Caleb24}
{Caleb} M.,  et~al., 2024, \mn@doi [Nature Astronomy] {10.1038/s41550-024-02277-w}, \href {https://ui.adsabs.harvard.edu/abs/2024NatAs...8.1159C} {8, 1159}

\bibitem[\protect\citeauthoryear{{Chen}, {Barr}, {Karuppusamy}, {Kramer}  \& {Stappers}}{{Chen} et~al.}{2021}]{Mosaic}
{Chen} W.,  {Barr} E.,  {Karuppusamy} R.,  {Kramer} M.,   {Stappers} B.,  2021, \mn@doi [Journal of Astronomical Instrumentation] {10.1142/S2251171721500136}, \href {https://ui.adsabs.harvard.edu/abs/2021JAI....1050013C} {10, 2150013}

\bibitem[\protect\citeauthoryear{{Chen} et~al.,}{{Chen} et~al.}{2022}]{RRAT_micro}
{Chen} J.~L.,  et~al., 2022, \mn@doi [\apj] {10.3847/1538-4357/ac75d1}, \href {https://ui.adsabs.harvard.edu/abs/2022ApJ...934...24C} {934, 24}

\bibitem[\protect\citeauthoryear{Clark, La~Plante  \& Greenhill}{Clark et~al.}{2011}]{clark_accelerating_2011}
Clark M.~A.,  La~Plante P.~C.,   Greenhill L.~J.,  2011, Accelerating {Radio} {Astronomy} {Cross}-{Correlation} with {Graphics} {Processing} {Units}, \mn@doi{10.48550/arXiv.1107.4264}, \url {https://ui.adsabs.harvard.edu/abs/2011arXiv1107.4264C}

\bibitem[\protect\citeauthoryear{{Clark} et~al.,}{{Clark} et~al.}{2023}]{Clark23}
{Clark} C.~J.,  et~al., 2023, \mn@doi [\mnras] {10.1093/mnras/stac3742}, \href {https://ui.adsabs.harvard.edu/abs/2023MNRAS.519.5590C} {519, 5590}

\bibitem[\protect\citeauthoryear{{Cordes} \& {Lazio}}{{Cordes} \& {Lazio}}{2002}]{Cordes02}
{Cordes} J.~M.,  {Lazio} T.~J.~W.,  2002, arXiv e-prints, \href {https://ui.adsabs.harvard.edu/abs/2002astro.ph..7156C} {pp astro--ph/0207156}

\bibitem[\protect\citeauthoryear{{Cordes} \& {McLaughlin}}{{Cordes} \& {McLaughlin}}{2003}]{FRT_cordes}
{Cordes} J.~M.,  {McLaughlin} M.~A.,  2003, \mn@doi [\apj] {10.1086/378231}, \href {https://ui.adsabs.harvard.edu/abs/2003ApJ...596.1142C} {596, 1142}

\bibitem[\protect\citeauthoryear{{Cordes} et~al.,}{{Cordes} et~al.}{2006}]{Cordes06}
{Cordes} J.~M.,  et~al., 2006, \mn@doi [\apj] {10.1086/498335}, \href {https://ui.adsabs.harvard.edu/abs/2006ApJ...637..446C} {637, 446}

\bibitem[\protect\citeauthoryear{{Cotton}, {Kothes}, {Camilo}, {Chandra}, {Buchner}  \& {Nyamai}}{{Cotton} et~al.}{2024}]{Cotton24}
{Cotton} W.~D.,  {Kothes} R.,  {Camilo} F.,  {Chandra} P.,  {Buchner} S.,   {Nyamai} M.,  2024, \mn@doi [\apjs] {10.3847/1538-4365/ad0ecb}, \href {https://ui.adsabs.harvard.edu/abs/2024ApJS..270...21C} {270, 21}

\bibitem[\protect\citeauthoryear{{Dang} et~al.,}{{Dang} et~al.}{2024}]{RRAT_RVM}
{Dang} S.~J.,  et~al., 2024, \mn@doi [\mnras] {10.1093/mnras/stae046}, \href {https://ui.adsabs.harvard.edu/abs/2024MNRAS.528.1213D} {528, 1213}

\bibitem[\protect\citeauthoryear{Deller, Tingay, Bailes  \& West}{Deller et~al.}{2007}]{deller_difx_2007}
Deller A.~T.,  Tingay S.~J.,  Bailes M.,   West C.,  2007, \mn@doi [Publications of the Astronomical Society of the Pacific] {10.1086/513572}, 119, 318

\bibitem[\protect\citeauthoryear{Deller et~al.,}{Deller et~al.}{2011}]{deller_difx-2_2011}
Deller A.~T.,  et~al., 2011, \mn@doi [Publications of the Astronomical Society of the Pacific] {10.1086/658907}, 123, 275

\bibitem[\protect\citeauthoryear{{Dong} et~al.,}{{Dong} et~al.}{2023}]{Dong23}
{Dong} F.~A.,  et~al., 2023, \mn@doi [\mnras] {10.1093/mnras/stad2012}, \href {https://ui.adsabs.harvard.edu/abs/2023MNRAS.524.5132D} {524, 5132}

\bibitem[\protect\citeauthoryear{{Dong} et~al.,}{{Dong} et~al.}{2025a}]{Dong25b}
{Dong} F.~A.,  et~al., 2025a, \mn@doi [\apjl] {10.3847/2041-8213/adeaab}, \href {https://ui.adsabs.harvard.edu/abs/2025ApJ...988L..29D} {988, L29}

\bibitem[\protect\citeauthoryear{{Dong} et~al.,}{{Dong} et~al.}{2025b}]{Dong25}
{Dong} F.~A.,  et~al., 2025b, \mn@doi [\apjl] {10.3847/2041-8213/adfa8e}, \href {https://ui.adsabs.harvard.edu/abs/2025ApJ...990L..49D} {990, L49}

\bibitem[\protect\citeauthoryear{{Driessen} et~al.,}{{Driessen} et~al.}{2022}]{Driessen22}
{Driessen} L.~N.,  et~al., 2022, \mn@doi [\mnras] {10.1093/mnras/stac756}, \href {https://ui.adsabs.harvard.edu/abs/2022MNRAS.512.5037D} {512, 5037}

\bibitem[\protect\citeauthoryear{{Driessen} et~al.,}{{Driessen} et~al.}{2024}]{Driessen24}
{Driessen} L.~N.,  et~al., 2024, \mn@doi [\mnras] {10.1093/mnras/stad3329}, \href {https://ui.adsabs.harvard.edu/abs/2024MNRAS.527.3659D} {527, 3659}

\bibitem[\protect\citeauthoryear{{Eatough}, {Keane}  \& {Lyne}}{{Eatough} et~al.}{2009}]{Eatough09}
{Eatough} R.~P.,  {Keane} E.~F.,   {Lyne} A.~G.,  2009, \mn@doi [\mnras] {10.1111/j.1365-2966.2009.14524.x}, \href {https://ui.adsabs.harvard.edu/abs/2009MNRAS.395..410E} {395, 410}

\bibitem[\protect\citeauthoryear{{Frail}, {Goss}  \& {Whiteoak}}{{Frail} et~al.}{1994}]{Frail94}
{Frail} D.~A.,  {Goss} W.~M.,   {Whiteoak} J.~B.~Z.,  1994, \mn@doi [\apj] {10.1086/175038}, \href {https://ui.adsabs.harvard.edu/abs/1994ApJ...437..781F} {437, 781}

\bibitem[\protect\citeauthoryear{{Good} et~al.,}{{Good} et~al.}{2021}]{Good21}
{Good} D.~C.,  et~al., 2021, \mn@doi [\apj] {10.3847/1538-4357/ac1da6}, \href {https://ui.adsabs.harvard.edu/abs/2021ApJ...922...43G} {922, 43}

\bibitem[\protect\citeauthoryear{{Grover}, {Bhat}  \& {McSweeney}}{{Grover} et~al.}{2024}]{grover_FFA}
{Grover} G.,  {Bhat} R.,   {McSweeney} S.,  2024, \mn@doi [\pasa] {10.1017/pasa.2024.67}, \href {https://ui.adsabs.harvard.edu/abs/2024PASA...41...46G} {41, e046}

\bibitem[\protect\citeauthoryear{Gupta}{Gupta}{2022}]{Gupta2022}
Gupta V.,  2022, Phd thesis, Swinburne University of Technology, Melbourne, Victoria

\bibitem[\protect\citeauthoryear{{Han} et~al.,}{{Han} et~al.}{2021}]{Han21}
{Han} J.~L.,  et~al., 2021, \mn@doi [Research in Astronomy and Astrophysics] {10.1088/1674-4527/21/5/107}, \href {https://ui.adsabs.harvard.edu/abs/2021RAA....21..107H} {21, 107}

\bibitem[\protect\citeauthoryear{{Han} et~al.,}{{Han} et~al.}{2025}]{Han25}
{Han} J.~L.,  et~al., 2025, \mn@doi [Research in Astronomy and Astrophysics] {10.1088/1674-4527/ada3b7}, \href {https://ui.adsabs.harvard.edu/abs/2025RAA....25a4001H} {25, 014001}

\bibitem[\protect\citeauthoryear{{Hankins}}{{Hankins}}{1996}]{Hankin_review}
{Hankins} T.~H.,  1996, in {Johnston} S.,  {Walker} M.~A.,   {Bailes} M.,  eds,  Astronomical Society of the Pacific Conference Series Vol. 105, IAU Colloq. 160: Pulsars: Problems and Progress. p.~197

\bibitem[\protect\citeauthoryear{{Hessels}, {Ransom}, {Kaspi}, {Roberts}, {Champion}  \& {Stappers}}{{Hessels} et~al.}{2008}]{Hessels08}
{Hessels} J.~W.~T.,  {Ransom} S.~M.,  {Kaspi} V.~M.,  {Roberts} M.~S.~E.,  {Champion} D.~J.,   {Stappers} B.~W.,  2008, in {Bassa} C.,  {Wang} Z.,  {Cumming} A.,   {Kaspi} V.~M.,  eds,  American Institute of Physics Conference Series Vol. 983, 40 Years of Pulsars: Millisecond Pulsars, Magnetars and More. AIP, pp 613--615 (\mn@eprint {arXiv} {0710.1745}), \mn@doi{10.1063/1.2900310}

\bibitem[\protect\citeauthoryear{{Hobbs}, {Edwards}  \& {Manchester}}{{Hobbs} et~al.}{2006}]{Hobbs06}
{Hobbs} G.~B.,  {Edwards} R.~T.,   {Manchester} R.~N.,  2006, \mn@doi [\mnras] {10.1111/j.1365-2966.2006.10302.x}, \href {https://ui.adsabs.harvard.edu/abs/2006MNRAS.369..655H} {369, 655}

\bibitem[\protect\citeauthoryear{{Hurley-Walker} et~al.,}{{Hurley-Walker} et~al.}{2022}]{Hurley22}
{Hurley-Walker} N.,  et~al., 2022, \mn@doi [\nat] {10.1038/s41586-021-04272-x}, \href {https://ui.adsabs.harvard.edu/abs/2022Natur.601..526H} {601, 526}

\bibitem[\protect\citeauthoryear{{Hurley-Walker} et~al.,}{{Hurley-Walker} et~al.}{2023}]{18m_source}
{Hurley-Walker} N.,  et~al., 2023, \mn@doi [\nat] {10.1038/s41586-023-06202-5}, \href {https://ui.adsabs.harvard.edu/abs/2023Natur.619..487H} {619, 487}

\bibitem[\protect\citeauthoryear{{Hurley-Walker} et~al.,}{{Hurley-Walker} et~al.}{2024}]{Hurley24}
{Hurley-Walker} N.,  et~al., 2024, \mn@doi [\apjl] {10.3847/2041-8213/ad890e}, \href {https://ui.adsabs.harvard.edu/abs/2024ApJ...976L..21H} {976, L21}

\bibitem[\protect\citeauthoryear{{Hyman}, {Lazio}, {Kassim}, {Ray}, {Markwardt}  \& {Yusef-Zadeh}}{{Hyman} et~al.}{2005}]{Hyman05}
{Hyman} S.~D.,  {Lazio} T. J.~W.,  {Kassim} N.~E.,  {Ray} P.~S.,  {Markwardt} C.~B.,   {Yusef-Zadeh} F.,  2005, \mn@doi [\nat] {10.1038/nature03400}, \href {https://ui.adsabs.harvard.edu/abs/2005Natur.434...50H} {434, 50}

\bibitem[\protect\citeauthoryear{{Jankowski} et~al.,}{{Jankowski} et~al.}{2023}]{Jankowski23}
{Jankowski} F.,  et~al., 2023, \mn@doi [\mnras] {10.1093/mnras/stad2041}, \href {https://ui.adsabs.harvard.edu/abs/2023MNRAS.524.4275J} {524, 4275}

\bibitem[\protect\citeauthoryear{{Johnston} \& {Karastergiou}}{{Johnston} \& {Karastergiou}}{2019}]{Johnston19}
{Johnston} S.,  {Karastergiou} A.,  2019, \mn@doi [\mnras] {10.1093/mnras/stz400}, \href {https://ui.adsabs.harvard.edu/abs/2019MNRAS.485..640J} {485, 640}

\bibitem[\protect\citeauthoryear{{Jonas} \& {MeerKAT Team}}{{Jonas} \& {MeerKAT Team}}{2016}]{Jonas16}
{Jonas} J.,  {MeerKAT Team} 2016, in MeerKAT Science: On the Pathway to the SKA. p.~1, \mn@doi{10.22323/1.277.0001}

\bibitem[\protect\citeauthoryear{{Kaspi} \& {Beloborodov}}{{Kaspi} \& {Beloborodov}}{2017}]{magnetar_review_kaspi}
{Kaspi} V.~M.,  {Beloborodov} A.~M.,  2017, \mn@doi [\araa] {10.1146/annurev-astro-081915-023329}, \href {https://ui.adsabs.harvard.edu/abs/2017ARA&A..55..261K} {55, 261}

\bibitem[\protect\citeauthoryear{{Keane} \& {McLaughlin}}{{Keane} \& {McLaughlin}}{2011}]{RRAT_review}
{Keane} E.~F.,  {McLaughlin} M.~A.,  2011, \mn@doi [Bulletin of the Astronomical Society of India] {10.48550/arXiv.1109.6896}, \href {https://ui.adsabs.harvard.edu/abs/2011BASI...39..333K} {39, 333}

\bibitem[\protect\citeauthoryear{{Keane} et~al.,}{{Keane} et~al.}{2018}]{Keane18}
{Keane} E.~F.,  et~al., 2018, \mn@doi [\mnras] {10.1093/mnras/stx2126}, \href {https://ui.adsabs.harvard.edu/abs/2018MNRAS.473..116K} {473, 116}

\bibitem[\protect\citeauthoryear{{Keith} et~al.,}{{Keith} et~al.}{2010}]{Keith10}
{Keith} M.~J.,  et~al., 2010, \mn@doi [\mnras] {10.1111/j.1365-2966.2010.17325.x}, \href {https://ui.adsabs.harvard.edu/abs/2010MNRAS.409..619K} {409, 619}

\bibitem[\protect\citeauthoryear{{Kramer}, {Liu}, {Desvignes}, {Karuppusamy}  \& {Stappers}}{{Kramer} et~al.}{2024}]{microstrucutre_kramer}
{Kramer} M.,  {Liu} K.,  {Desvignes} G.,  {Karuppusamy} R.,   {Stappers} B.~W.,  2024, \mn@doi [Nature Astronomy] {10.1038/s41550-023-02125-3}, \href {https://ui.adsabs.harvard.edu/abs/2024NatAs...8..230K} {8, 230}

\bibitem[\protect\citeauthoryear{{Law} et~al.,}{{Law} et~al.}{2018}]{Law18}
{Law} C.~J.,  et~al., 2018, \mn@doi [\apjs] {10.3847/1538-4365/aab77b}, \href {https://ui.adsabs.harvard.edu/abs/2018ApJS..236....8L} {236, 8}

\bibitem[\protect\citeauthoryear{{Lee} et~al.,}{{Lee} et~al.}{2025}]{Lee25}
{Lee} Y.~W.~J.,  et~al., 2025, \mn@doi [Nature Astronomy] {10.1038/s41550-024-02452-z}, \href {https://ui.adsabs.harvard.edu/abs/2025NatAs...9..393L} {9, 393}

\bibitem[\protect\citeauthoryear{Lehmensiek \& Theron}{Lehmensiek \& Theron}{2012}]{L_band}
Lehmensiek R.,  Theron I.~P.,  2012, in 2012 International Conference on Electromagnetics in Advanced Applications. pp 321--324, \mn@doi{10.1109/ICEAA.2012.6328642}

\bibitem[\protect\citeauthoryear{Lehmensiek \& Theron}{Lehmensiek \& Theron}{2014}]{UHF_band}
Lehmensiek R.,  Theron I.~P.,  2014, in The 8th European Conference on Antennas and Propagation (EuCAP 2014). pp 880--884, \mn@doi{10.1109/EuCAP.2014.6901903}

\bibitem[\protect\citeauthoryear{{Liu} et~al.,}{{Liu} et~al.}{2025}]{Liu25}
{Liu} X.,  et~al., 2025, \mn@doi [\apj] {10.3847/1538-4357/ade689}, \href {https://ui.adsabs.harvard.edu/abs/2025ApJ...988..175L} {988, 175}

\bibitem[\protect\citeauthoryear{{Lorimer}, {McLaughlin}  \& {Bailes}}{{Lorimer} et~al.}{2024}]{Lorimer24}
{Lorimer} D.~R.,  {McLaughlin} M.~A.,   {Bailes} M.,  2024, \mn@doi [\apss] {10.1007/s10509-024-04322-6}, \href {https://ui.adsabs.harvard.edu/abs/2024Ap&SS.369...59L} {369, 59}

\bibitem[\protect\citeauthoryear{{Lu} et~al.,}{{Lu} et~al.}{2019}]{RRAT_modes}
{Lu} J.,  et~al., 2019, \mn@doi [Science China Physics, Mechanics, and Astronomy] {10.1007/s11433-018-9372-7}, \href {https://ui.adsabs.harvard.edu/abs/2019SCPMA..6259503L} {62, 959503}

\bibitem[\protect\citeauthoryear{{Lyne}, {McLaughlin}, {Keane}, {Kramer}, {Espinoza}, {Stappers}, {Palliyaguru}  \& {Miller}}{{Lyne} et~al.}{2009}]{Lyne09}
{Lyne} A.~G.,  {McLaughlin} M.~A.,  {Keane} E.~F.,  {Kramer} M.,  {Espinoza} C.~M.,  {Stappers} B.~W.,  {Palliyaguru} N.~T.,   {Miller} J.,  2009, \mn@doi [\mnras] {10.1111/j.1365-2966.2009.15668.x}, \href {https://ui.adsabs.harvard.edu/abs/2009MNRAS.400.1439L} {400, 1439}

\bibitem[\protect\citeauthoryear{{Maciesiak}, {Gil}  \& {Melikidze}}{{Maciesiak} et~al.}{2012}]{Maciesiak12}
{Maciesiak} K.,  {Gil} J.,   {Melikidze} G.,  2012, \mn@doi [\mnras] {10.1111/j.1365-2966.2012.21246.x}, \href {https://ui.adsabs.harvard.edu/abs/2012MNRAS.424.1762M} {424, 1762}

\bibitem[\protect\citeauthoryear{{Macquart} et~al.,}{{Macquart} et~al.}{2010}]{FRT_and_extremephysics}
{Macquart} J.-P.,  et~al., 2010, \mn@doi [\pasa] {10.1071/AS09082}, \href {https://ui.adsabs.harvard.edu/abs/2010PASA...27..272M} {27, 272}

\bibitem[\protect\citeauthoryear{{Majid} et~al.,}{{Majid} et~al.}{2021}]{Majid21}
{Majid} W.~A.,  et~al., 2021, \mn@doi [\apjl] {10.3847/2041-8213/ac1921}, \href {https://ui.adsabs.harvard.edu/abs/2021ApJ...919L...6M} {919, L6}

\bibitem[\protect\citeauthoryear{{Manchester} et~al.,}{{Manchester} et~al.}{2001}]{Manchester01}
{Manchester} R.~N.,  et~al., 2001, \mn@doi [\mnras] {10.1046/j.1365-8711.2001.04751.x}, \href {https://ui.adsabs.harvard.edu/abs/2001MNRAS.328...17M} {328, 17}

\bibitem[\protect\citeauthoryear{{Mauch} et~al.,}{{Mauch} et~al.}{2020}]{Mauch20}
{Mauch} T.,  et~al., 2020, \mn@doi [\apj] {10.3847/1538-4357/ab5d2d}, \href {https://ui.adsabs.harvard.edu/abs/2020ApJ...888...61M} {888, 61}

\bibitem[\protect\citeauthoryear{{McConnell} et~al.,}{{McConnell} et~al.}{2020}]{McConnell20}
{McConnell} D.,  et~al., 2020, \mn@doi [\pasa] {10.1017/pasa.2020.41}, \href {https://ui.adsabs.harvard.edu/abs/2020PASA...37...48M} {37, e048}

\bibitem[\protect\citeauthoryear{{McLaughlin} \& {Cordes}}{{McLaughlin} \& {Cordes}}{2003}]{SP_vs_Periodicity_Maura}
{McLaughlin} M.~A.,  {Cordes} J.~M.,  2003, \mn@doi [\apj] {10.1086/378232}, \href {https://ui.adsabs.harvard.edu/abs/2003ApJ...596..982M} {596, 982}

\bibitem[\protect\citeauthoryear{{McLaughlin} et~al.,}{{McLaughlin} et~al.}{2006}]{McLaughlin06}
{McLaughlin} M.~A.,  et~al., 2006, \mn@doi [\nat] {10.1038/nature04440}, \href {https://ui.adsabs.harvard.edu/abs/2006Natur.439..817M} {439, 817}

\bibitem[\protect\citeauthoryear{{McLaughlin} et~al.,}{{McLaughlin} et~al.}{2009}]{McLaughlin09}
{McLaughlin} M.~A.,  et~al., 2009, \mn@doi [\mnras] {10.1111/j.1365-2966.2009.15584.x}, \href {https://ui.adsabs.harvard.edu/abs/2009MNRAS.400.1431M} {400, 1431}

\bibitem[\protect\citeauthoryear{{McSweeney} et~al.,}{{McSweeney} et~al.}{2025}]{McSweeney25b}
{McSweeney} S.~J.,  et~al., 2025, \mn@doi [\mnras] {10.1093/mnras/staf1203}, \href {https://ui.adsabs.harvard.edu/abs/2025MNRAS.542..203M} {542, 203}

\bibitem[\protect\citeauthoryear{{Mckinven} et~al.,}{{Mckinven} et~al.}{2025}]{Mckinven25}
{Mckinven} R.,  et~al., 2025, \mn@doi [\nat] {10.1038/s41586-024-08184-4}, \href {https://ui.adsabs.harvard.edu/abs/2025Natur.637...43M} {637, 43}

\bibitem[\protect\citeauthoryear{{Mcsweeney}, {Moseley}, {Hurley-Walker}, {Grover}, {Horv{\'a}th}, {Galvin}, {Meyers}  \& {Tan}}{{Mcsweeney} et~al.}{2025}]{Mcsweeney25}
{Mcsweeney} S.~J.,  {Moseley} J.,  {Hurley-Walker} N.,  {Grover} G.,  {Horv{\'a}th} C.,  {Galvin} T.~J.,  {Meyers} B.~W.,   {Tan} C.~M.,  2025, \mn@doi [\apj] {10.3847/1538-4357/adb27f}, \href {https://ui.adsabs.harvard.edu/abs/2025ApJ...981..143M} {981, 143}

\bibitem[\protect\citeauthoryear{{Men} \& {Barr}}{{Men} \& {Barr}}{2024}]{TransientX}
{Men} Y.,  {Barr} E.,  2024, \mn@doi [\aap] {10.1051/0004-6361/202348247}, \href {https://ui.adsabs.harvard.edu/abs/2024A&A...683A.183M} {683, A183}

\bibitem[\protect\citeauthoryear{{Mereghetti}, {Pons}  \& {Melatos}}{{Mereghetti} et~al.}{2015}]{magnetar_Review_mereghetti}
{Mereghetti} S.,  {Pons} J.~A.,   {Melatos} A.,  2015, \mn@doi [\ssr] {10.1007/s11214-015-0146-y}, \href {https://ui.adsabs.harvard.edu/abs/2015SSRv..191..315M} {191, 315}

\bibitem[\protect\citeauthoryear{{Meyers} et~al.,}{{Meyers} et~al.}{2018}]{Meyers18}
{Meyers} B.~W.,  et~al., 2018, \mn@doi [\apj] {10.3847/1538-4357/aaee7b}, \href {https://ui.adsabs.harvard.edu/abs/2018ApJ...869..134M} {869, 134}

\bibitem[\protect\citeauthoryear{{Meyers} et~al.,}{{Meyers} et~al.}{2019}]{Meyers19}
{Meyers} B.~W.,  et~al., 2019, \mn@doi [\pasa] {10.1017/pasa.2019.30}, \href {https://ui.adsabs.harvard.edu/abs/2019PASA...36...34M} {36, e034}

\bibitem[\protect\citeauthoryear{{Mitra}, {Arjunwadkar}  \& {Rankin}}{{Mitra} et~al.}{2015}]{Mitra_micro}
{Mitra} D.,  {Arjunwadkar} M.,   {Rankin} J.~M.,  2015, \mn@doi [\apj] {10.1088/0004-637X/806/2/236}, \href {https://ui.adsabs.harvard.edu/abs/2015ApJ...806..236M} {806, 236}

\bibitem[\protect\citeauthoryear{{Morello} et~al.,}{{Morello} et~al.}{2020a}]{12s_pulsar}
{Morello} V.,  et~al., 2020a, \mn@doi [\mnras] {10.1093/mnras/staa321}, \href {https://ui.adsabs.harvard.edu/abs/2020MNRAS.493.1165M} {493, 1165}

\bibitem[\protect\citeauthoryear{{Morello}, {Barr}, {Stappers}, {Keane}  \& {Lyne}}{{Morello} et~al.}{2020b}]{FFA_morello}
{Morello} V.,  {Barr} E.~D.,  {Stappers} B.~W.,  {Keane} E.~F.,   {Lyne} A.~G.,  2020b, \mn@doi [\mnras] {10.1093/mnras/staa2291}, \href {https://ui.adsabs.harvard.edu/abs/2020MNRAS.497.4654M} {497, 4654}

\bibitem[\protect\citeauthoryear{{Morello}, {Rajwade}  \& {Stappers}}{{Morello} et~al.}{2022}]{Morello22}
{Morello} V.,  {Rajwade} K.~M.,   {Stappers} B.~W.,  2022, \mn@doi [\mnras] {10.1093/mnras/stab3493}, \href {https://ui.adsabs.harvard.edu/abs/2022MNRAS.510.1393M} {510, 1393}

\bibitem[\protect\citeauthoryear{{Offringa} et~al.,}{{Offringa} et~al.}{2014}]{Offringa14}
{Offringa} A.~R.,  et~al., 2014, \mn@doi [\mnras] {10.1093/mnras/stu1368}, \href {https://ui.adsabs.harvard.edu/abs/2014MNRAS.444..606O} {444, 606}

\bibitem[\protect\citeauthoryear{{Padmanabh} et~al.,}{{Padmanabh} et~al.}{2023}]{Padmanabh23}
{Padmanabh} P.~V.,  et~al., 2023, \mn@doi [\mnras] {10.1093/mnras/stad1900}, \href {https://ui.adsabs.harvard.edu/abs/2023MNRAS.524.1291P} {524, 1291}

\bibitem[\protect\citeauthoryear{{Panda}, {Bhattacharyya}, {Dudeja}, {Kudale}  \& {Roy}}{{Panda} et~al.}{2024}]{Panda24}
{Panda} U.,  {Bhattacharyya} S.,  {Dudeja} C.,  {Kudale} S.,   {Roy} J.,  2024, The Astronomer's Telegram, \href {https://ui.adsabs.harvard.edu/abs/2024ATel16494....1P} {16494, 1}

\bibitem[\protect\citeauthoryear{{Pastor-Marazuela} et~al.,}{{Pastor-Marazuela} et~al.}{2023}]{Pastor-Marazuela23}
{Pastor-Marazuela} I.,  et~al., 2023, \mn@doi [\aap] {10.1051/0004-6361/202243339}, \href {https://ui.adsabs.harvard.edu/abs/2023A&A...678A.149P} {678, A149}

\bibitem[\protect\citeauthoryear{Pastor-Marazuela et~al.,}{Pastor-Marazuela et~al.}{2025}]{pastor-marazuela_localisation_2025}
Pastor-Marazuela I.,  et~al., 2025, Localisation and host galaxy identification of new {Fast} {Radio} {Bursts} with {MeerKAT}, \mn@doi{10.48550/arXiv.2507.05982}, \url {http://arxiv.org/abs/2507.05982}

\bibitem[\protect\citeauthoryear{{Patel} et~al.,}{{Patel} et~al.}{2018}]{Patel18}
{Patel} C.,  et~al., 2018, \mn@doi [\apj] {10.3847/1538-4357/aaee65}, \href {https://ui.adsabs.harvard.edu/abs/2018ApJ...869..181P} {869, 181}

\bibitem[\protect\citeauthoryear{{Platts}, {Weltman}, {Walters}, {Tendulkar}, {Gordin}  \& {Kandhai}}{{Platts} et~al.}{2019}]{Platts19}
{Platts} E.,  {Weltman} A.,  {Walters} A.,  {Tendulkar} S.~P.,  {Gordin} J.~E.~B.,   {Kandhai} S.,  2019, \mn@doi [\physrep] {10.1016/j.physrep.2019.06.003}, \href {https://ui.adsabs.harvard.edu/abs/2019PhR...821....1P} {821, 1}

\bibitem[\protect\citeauthoryear{{Posselt} et~al.,}{{Posselt} et~al.}{2021}]{Posselt21}
{Posselt} B.,  et~al., 2021, \mn@doi [\mnras] {10.1093/mnras/stab2775}, \href {https://ui.adsabs.harvard.edu/abs/2021MNRAS.508.4249P} {508, 4249}

\bibitem[\protect\citeauthoryear{{Price}, {Flynn}  \& {Deller}}{{Price} et~al.}{2021}]{Price21}
{Price} D.~C.,  {Flynn} C.,   {Deller} A.,  2021, \mn@doi [\pasa] {10.1017/pasa.2021.33}, \href {https://ui.adsabs.harvard.edu/abs/2021PASA...38...38P} {38, e038}

\bibitem[\protect\citeauthoryear{{Rajwade} \& {van Leeuwen}}{{Rajwade} \& {van Leeuwen}}{2024}]{FRBsearch_review}
{Rajwade} K.~M.,  {van Leeuwen} J.,  2024, \mn@doi [Universe] {10.3390/universe10040158}, \href {https://ui.adsabs.harvard.edu/abs/2024Univ...10..158R} {10, 158}

\bibitem[\protect\citeauthoryear{{Rajwade} et~al.,}{{Rajwade} et~al.}{2022}]{Rajwade22}
{Rajwade} K.~M.,  et~al., 2022, \mn@doi [\mnras] {10.1093/mnras/stac1450}, \href {https://ui.adsabs.harvard.edu/abs/2022MNRAS.514.1961R} {514, 1961}

\bibitem[\protect\citeauthoryear{{Rajwade} et~al.,}{{Rajwade} et~al.}{2024}]{Rajwade24}
{Rajwade} K.~M.,  et~al., 2024, \mn@doi [\mnras] {10.1093/mnras/stae1652}, \href {https://ui.adsabs.harvard.edu/abs/2024MNRAS.532.3881R} {532, 3881}

\bibitem[\protect\citeauthoryear{Ransom, Eikenberry  \& Middleditch}{Ransom et~al.}{2002}]{Ransom_2002}
Ransom S.~M.,  Eikenberry S.~S.,   Middleditch J.,  2002, \mn@doi [The Astronomical Journal] {10.1086/342285}, 124, 1788–1809

\bibitem[\protect\citeauthoryear{{Rea} et~al.,}{{Rea} et~al.}{2009}]{Rea09}
{Rea} N.,  et~al., 2009, \mn@doi [\apjl] {10.1088/0004-637X/703/1/L41}, \href {https://ui.adsabs.harvard.edu/abs/2009ApJ...703L..41R} {703, L41}

\bibitem[\protect\citeauthoryear{{Sanidas}, {Caleb}, {Driessen}, {Morello}, {Rajwade}  \& {Stappers}}{{Sanidas} et~al.}{2018}]{Sanidas18}
{Sanidas} S.,  {Caleb} M.,  {Driessen} L.,  {Morello} V.,  {Rajwade} K.,   {Stappers} B.~W.,  2018, in {Weltevrede} P.,  {Perera} B.~B.~P.,  {Preston} L.~L.,   {Sanidas} S.,  eds,  IAU Symposium Vol. 337, Pulsar Astrophysics the Next Fifty Years. pp 406--407, \mn@doi{10.1017/S1743921317009310}

\bibitem[\protect\citeauthoryear{{Sengar} et~al.,}{{Sengar} et~al.}{2023}]{Sengar23}
{Sengar} R.,  et~al., 2023, \mn@doi [\mnras] {10.1093/mnras/stad508}, \href {https://ui.adsabs.harvard.edu/abs/2023MNRAS.522.1071S} {522, 1071}

\bibitem[\protect\citeauthoryear{{Singh}, {Roy}, {Panda}, {Bhattacharyya}, {Morello}, {Stappers}, {Ray}  \& {McLaughlin}}{{Singh} et~al.}{2022}]{singh_FFA}
{Singh} S.,  {Roy} J.,  {Panda} U.,  {Bhattacharyya} B.,  {Morello} V.,  {Stappers} B.~W.,  {Ray} P.~S.,   {McLaughlin} M.~A.,  2022, \mn@doi [\apj] {10.3847/1538-4357/ac7b91}, \href {https://ui.adsabs.harvard.edu/abs/2022ApJ...934..138S} {934, 138}

\bibitem[\protect\citeauthoryear{{Singh}, {Roy}, {Sharma}, {Bhattacharyya}  \& {Kudale}}{{Singh} et~al.}{2023}]{GHRSS_VI}
{Singh} S.,  {Roy} J.,  {Sharma} S.~S.,  {Bhattacharyya} B.,   {Kudale} S.,  2023, \mn@doi [\apj] {10.3847/1538-4357/ace781}, \href {https://ui.adsabs.harvard.edu/abs/2023ApJ...954..160S} {954, 160}

\bibitem[\protect\citeauthoryear{{Skrzypczak}, {Basu}, {Mitra}, {Melikidze}, {Maciesiak}, {Koralewska}  \& {Filothodoros}}{{Skrzypczak} et~al.}{2018}]{Skrzypczak18}
{Skrzypczak} A.,  {Basu} R.,  {Mitra} D.,  {Melikidze} G.~I.,  {Maciesiak} K.,  {Koralewska} O.,   {Filothodoros} A.,  2018, \mn@doi [\apj] {10.3847/1538-4357/aaa758}, \href {https://ui.adsabs.harvard.edu/abs/2018ApJ...854..162S} {854, 162}

\bibitem[\protect\citeauthoryear{{Staelin}}{{Staelin}}{1969}]{staelin}
{Staelin} D.~H.,  1969, \mn@doi [IEEE Proceedings] {10.1109/PROC.1969.7051}, \href {https://ui.adsabs.harvard.edu/abs/1969IEEEP..57..724S} {57, 724}

\bibitem[\protect\citeauthoryear{{Stappers}}{{Stappers}}{2016}]{meertrap_concept}
{Stappers} B.,  2016, in MeerKAT Science: On the Pathway to the SKA. p.~10

\bibitem[\protect\citeauthoryear{{Stappers}, {Keane}, {Kramer}, {Possenti}  \& {Stairs}}{{Stappers} et~al.}{2018}]{Stappers18}
{Stappers} B.~W.,  {Keane} E.~F.,  {Kramer} M.,  {Possenti} A.,   {Stairs} I.~H.,  2018, \mn@doi [Philosophical Transactions of the Royal Society of London Series A] {10.1098/rsta.2017.0293}, \href {https://ui.adsabs.harvard.edu/abs/2018RSPTA.37670293S} {376, 20170293}

\bibitem[\protect\citeauthoryear{{Surnis} et~al.,}{{Surnis} et~al.}{2023}]{Surnis23}
{Surnis} M.~P.,  et~al., 2023, \mn@doi [\mnras] {10.1093/mnrasl/slad082}, \href {https://ui.adsabs.harvard.edu/abs/2023MNRAS.tmpL..84S} {}

\bibitem[\protect\citeauthoryear{{Tan} et~al.,}{{Tan} et~al.}{2018}]{23s_pulsar}
{Tan} C.~M.,  et~al., 2018, \mn@doi [\apj] {10.3847/1538-4357/aade88}, \href {https://ui.adsabs.harvard.edu/abs/2018ApJ...866...54T} {866, 54}

\bibitem[\protect\citeauthoryear{{Tian} et~al.,}{{Tian} et~al.}{2024}]{Tian24b}
{Tian} J.,  et~al., 2024, \mn@doi [\mnras] {10.1093/mnras/stae2013}, \href {https://ui.adsabs.harvard.edu/abs/2024MNRAS.533.3174T} {533, 3174}

\bibitem[\protect\citeauthoryear{{Turner} et~al.,}{{Turner} et~al.}{2025}]{MeerTRAP_james}
{Turner} J.~D.,  et~al., 2025, \mn@doi [\mnras] {10.1093/mnras/staf098}, \href {https://ui.adsabs.harvard.edu/abs/2025MNRAS.537.1070T} {537, 1070}

\bibitem[\protect\citeauthoryear{{Venkatraman Krishnan} et~al.,}{{Venkatraman Krishnan} et~al.}{2020}]{Venkatraman20}
{Venkatraman Krishnan} V.,  et~al., 2020, \mn@doi [\mnras] {10.1093/mnras/staa111}, \href {https://ui.adsabs.harvard.edu/abs/2020MNRAS.492.4752V} {492, 4752}

\bibitem[\protect\citeauthoryear{{Vink}}{{Vink}}{2020}]{Vink20}
{Vink} J.,  2020, {Physics and Evolution of Supernova Remnants}, \mn@doi{10.1007/978-3-030-55231-2.
}

\bibitem[\protect\citeauthoryear{{Wang} et~al.,}{{Wang} et~al.}{2024}]{Wang24}
{Wang} Z.,  et~al., 2024, \mn@doi [arXiv e-prints] {10.48550/arXiv.2411.16606}, \href {https://ui.adsabs.harvard.edu/abs/2024arXiv241116606W} {p. arXiv:2411.16606}

\bibitem[\protect\citeauthoryear{{Wang} et~al.,}{{Wang} et~al.}{2025a}]{WangZ25}
{Wang} Z.,  et~al., 2025a, \mn@doi [\pasa] {10.1017/pasa.2024.107}, \href {https://ui.adsabs.harvard.edu/abs/2025PASA...42....5W} {42, e005}

\bibitem[\protect\citeauthoryear{{Wang} et~al.,}{{Wang} et~al.}{2025b}]{WangY25}
{Wang} Y.,  et~al., 2025b, \mn@doi [\apjl] {10.3847/2041-8213/adbe61}, \href {https://ui.adsabs.harvard.edu/abs/2025ApJ...982L..53W} {982, L53}

\bibitem[\protect\citeauthoryear{{Weltevrede}}{{Weltevrede}}{2016}]{Weltevrede16}
{Weltevrede} P.,  2016, \mn@doi [\aap] {10.1051/0004-6361/201527950}, \href {https://ui.adsabs.harvard.edu/abs/2016A&A...590A.109W} {590, A109}

\bibitem[\protect\citeauthoryear{{Wenger} et~al.,}{{Wenger} et~al.}{2000}]{Wenger2000}
{Wenger} M.,  et~al., 2000, \mn@doi [\aaps] {10.1051/aas:2000332}, \href {https://ui.adsabs.harvard.edu/abs/2000A&AS..143....9W} {143, 9}

\bibitem[\protect\citeauthoryear{{Wielebinski}, {Junkes}  \& {Grahl}}{{Wielebinski} et~al.}{2011}]{Effelseberg_paper}
{Wielebinski} R.,  {Junkes} N.,   {Grahl} B.~H.,  2011, Journal of Astronomical History and Heritage, \href {https://ui.adsabs.harvard.edu/abs/2011JAHH...14....3W} {14, 3}

\bibitem[\protect\citeauthoryear{{Yamasaki} \& {Totani}}{{Yamasaki} \& {Totani}}{2020}]{Yamasaki20}
{Yamasaki} S.,  {Totani} T.,  2020, \mn@doi [\apj] {10.3847/1538-4357/ab58c4}, \href {https://ui.adsabs.harvard.edu/abs/2020ApJ...888..105Y} {888, 105}

\bibitem[\protect\citeauthoryear{{Yao}, {Manchester}  \& {Wang}}{{Yao} et~al.}{2017}]{YMW16}
{Yao} J.~M.,  {Manchester} R.~N.,   {Wang} N.,  2017, \mn@doi [\apj] {10.3847/1538-4357/835/1/29}, \href {https://ui.adsabs.harvard.edu/abs/2017ApJ...835...29Y} {835, 29}

\bibitem[\protect\citeauthoryear{{Zhang}}{{Zhang}}{2023}]{Zhang23}
{Zhang} B.,  2023, \mn@doi [Reviews of Modern Physics] {10.1103/RevModPhys.95.035005}, \href {https://ui.adsabs.harvard.edu/abs/2023RvMP...95c5005Z} {95, 035005}

\bibitem[\protect\citeauthoryear{{Zhang} et~al.,}{{Zhang} et~al.}{2024}]{Zhang24}
{Zhang} J.,  et~al., 2024, The Astronomer's Telegram, \href {https://ui.adsabs.harvard.edu/abs/2024ATel16433....1Z} {16433, 1}

\bibitem[\protect\citeauthoryear{{Zhou}, {Vink}, {Li}  \& {Dom{\v{c}}ek}}{{Zhou} et~al.}{2018}]{Zhou18}
{Zhou} P.,  {Vink} J.,  {Li} G.,   {Dom{\v{c}}ek} V.,  2018, \mn@doi [\apjl] {10.3847/2041-8213/aae07d}, \href {https://ui.adsabs.harvard.edu/abs/2018ApJ...865L...6Z} {865, L6}

\bibitem[\protect\citeauthoryear{{Zhou} et~al.,}{{Zhou} et~al.}{2023}]{Zhou23}
{Zhou} D.~J.,  et~al., 2023, \mn@doi [Research in Astronomy and Astrophysics] {10.1088/1674-4527/accc76}, \href {https://ui.adsabs.harvard.edu/abs/2023RAA....23j4001Z} {23, 104001}

\bibitem[\protect\citeauthoryear{{de Ruiter} et~al.,}{{de Ruiter} et~al.}{2025}]{Ruiter25}
{de Ruiter} I.,  et~al., 2025, \mn@doi [Nature Astronomy] {10.1038/s41550-025-02491-0}, \href {https://ui.adsabs.harvard.edu/abs/2025NatAs...9..672D} {9, 672}

\bibitem[\protect\citeauthoryear{{de Villiers}}{{de Villiers}}{2023}]{Villiers23}
{de Villiers} M.~S.,  2023, \mn@doi [\aj] {10.3847/1538-3881/acabc3}, \href {https://ui.adsabs.harvard.edu/abs/2023AJ....165...78D} {165, 78}

\bibitem[\protect\citeauthoryear{{van Heerden}, {Karastergiou}  \& {Roberts}}{{van Heerden} et~al.}{2017}]{heerden_2016}
{van Heerden} E.,  {Karastergiou} A.,   {Roberts} S.~J.,  2017, \mn@doi [\mnras] {10.1093/mnras/stw3068}, \href {https://ui.adsabs.harvard.edu/abs/2017MNRAS.467.1661V} {467, 1661}

\makeatother
\end{thebibliography}




\appendix
\section{Images}\label{append:images}
\begin{figure*}
\subfigure[MTP0055]{
  \label{fig:images_55}
  \includegraphics[width=.7\linewidth]{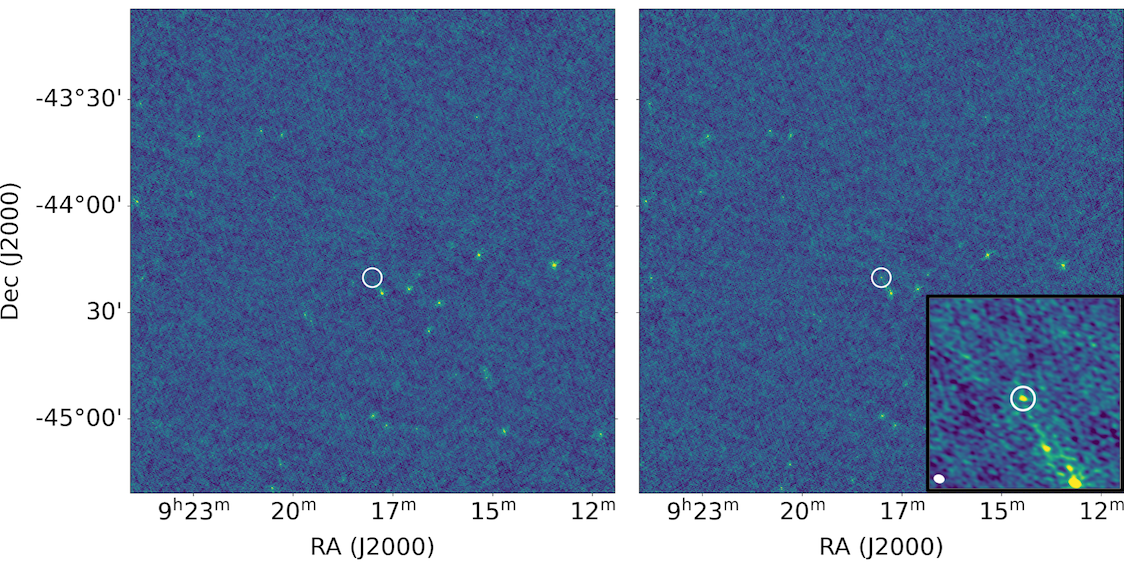}} \\
\subfigure[MTP0063]{
  \label{fig:images_63}
  \includegraphics[width=.7\linewidth]{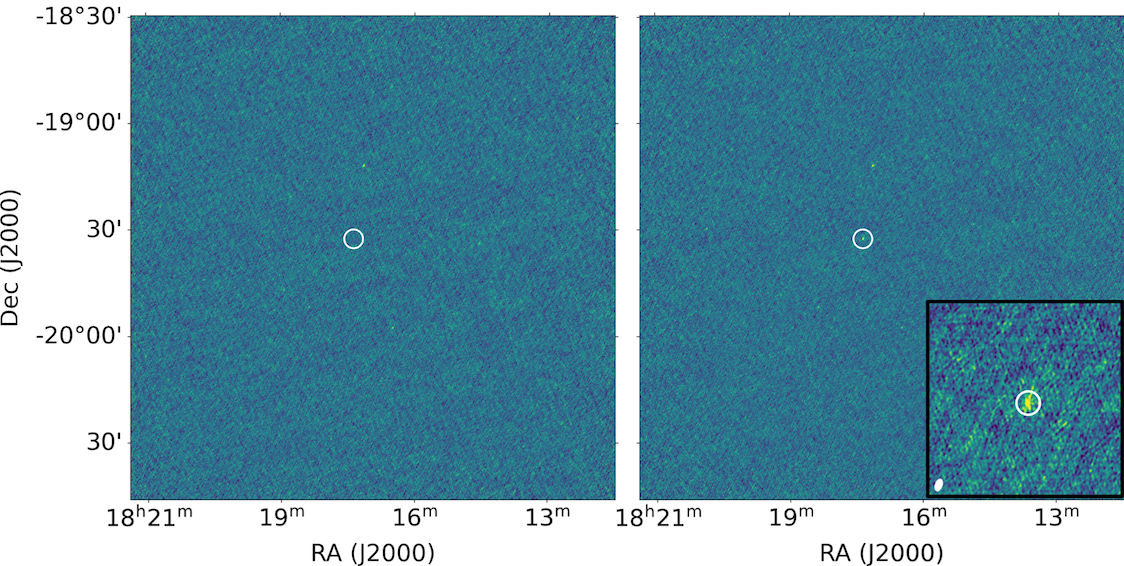}} \\
\subfigure[MTP0067]{
  \label{fig:images_67}
  \includegraphics[width=.7\linewidth]{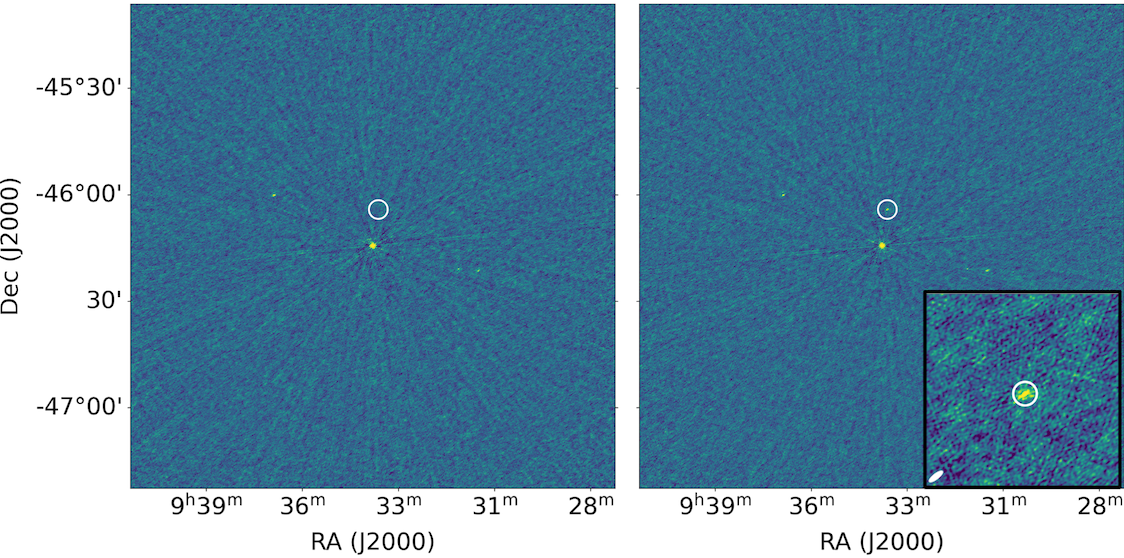}}
\caption{Images of the 9 MTP sources that triggered the transient buffer to save voltage data (see Section~\ref{sec:image}). For each source, we show the image integrated over the duration of the pulse (right) and before the pulse detection (left). The magenta circle in the on-pulse image marks the transient source identified at the pulse detection, and the inset at the bottom right corner is a zoomed-in view of the source. The synthesised beam is shown at the bottom left corner of the inset.}
\label{fig:images}
\end{figure*}

\renewcommand{\thefigure}{A\arabic{figure} (Continued.)}
\addtocounter{figure}{-1}

\begin{figure*}
\subfigure[MTP0069]{
  \label{fig:images_69}
  \includegraphics[width=.7\linewidth]{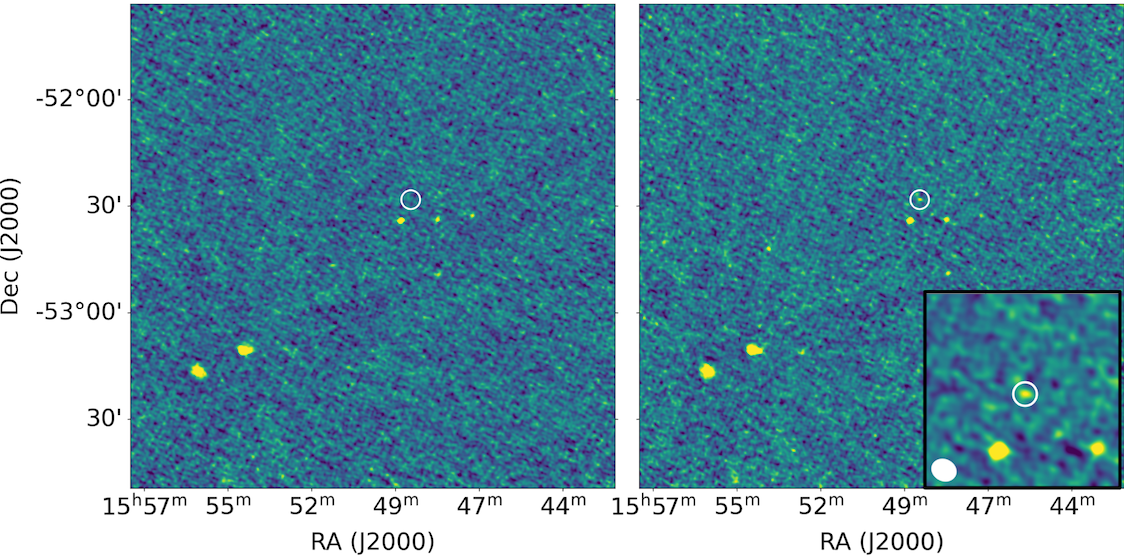}} \\
\subfigure[MTP0072]{
  \label{fig:images_72}
  \includegraphics[width=.7\linewidth]{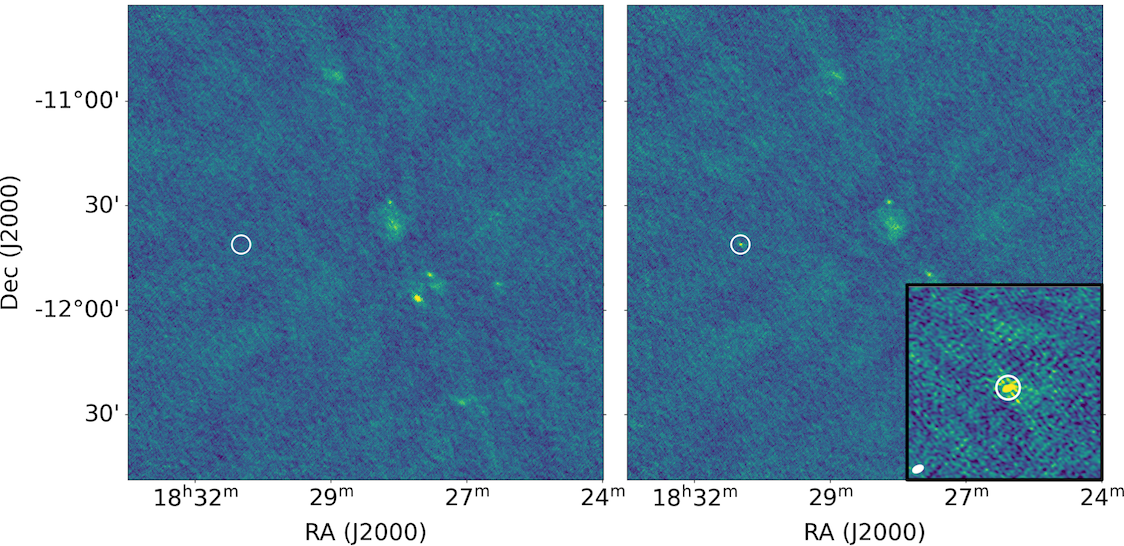}} \\
\subfigure[MTP0074]{
  \label{fig:images_74}
  \includegraphics[width=.7\linewidth]{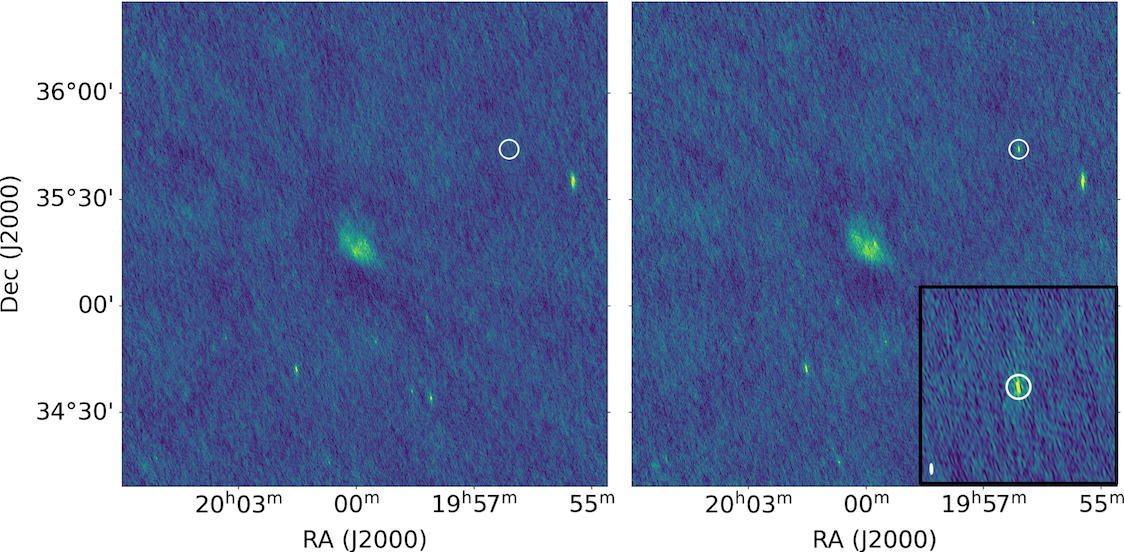}}
\caption{
}
\end{figure*}

\renewcommand{\thefigure}{A\arabic{figure} (Continued.)}
\addtocounter{figure}{-1}

\begin{figure*}
\subfigure[MTP0079]{
  \label{fig:images_79}
  \includegraphics[width=.7\linewidth]{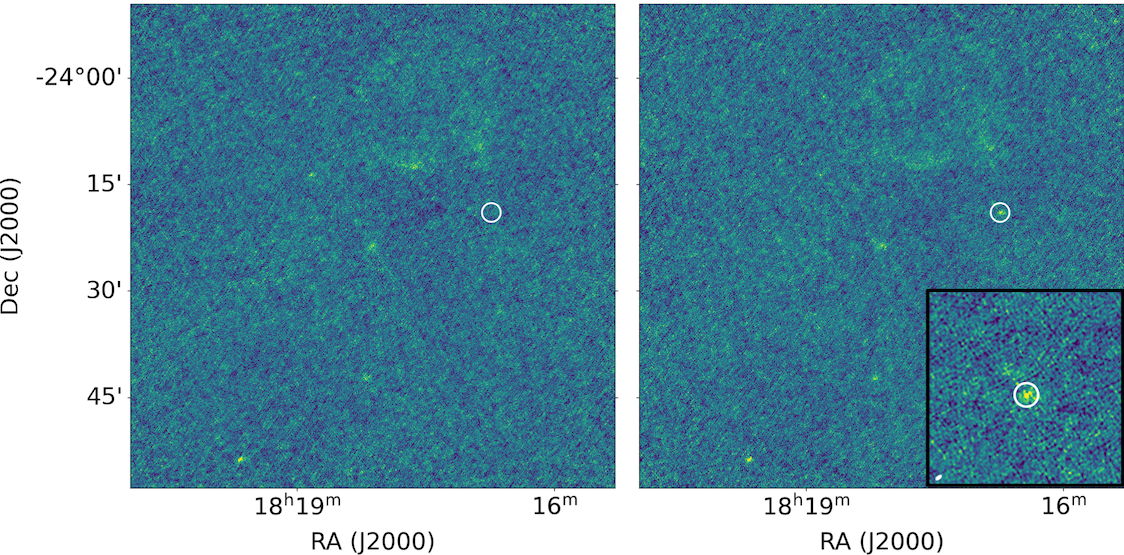}} \\
\subfigure[MTP0080]{
  \label{fig:images_80}
  \includegraphics[width=.7\linewidth]{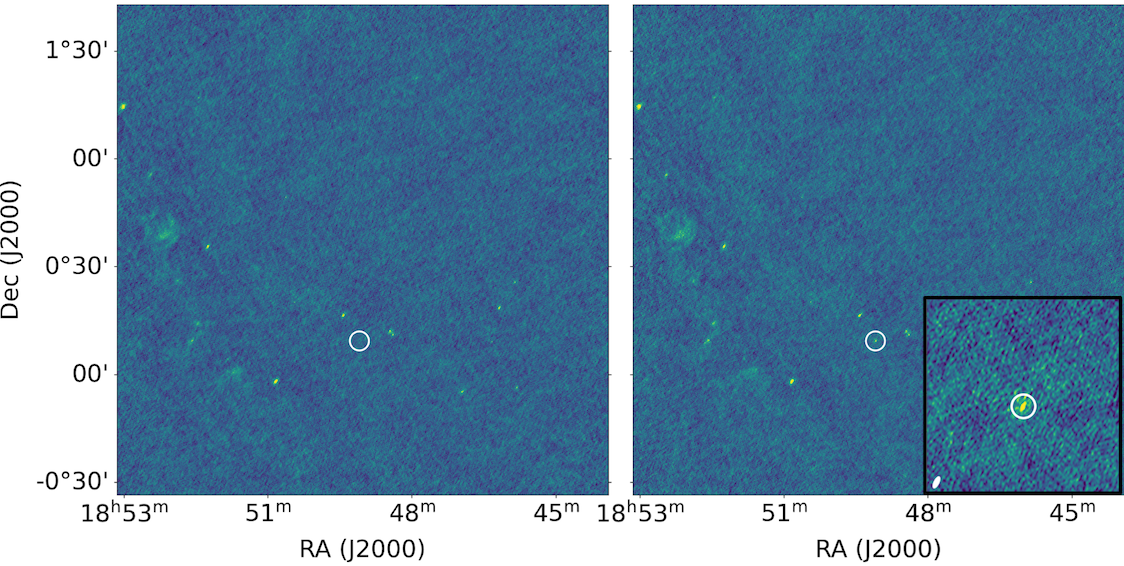}}
\caption{
}
\end{figure*}

\renewcommand{\thefigure}{A\arabic{figure}}


\bsp	
\label{lastpage}
\end{document}